\documentclass[reprint,aps,prd,longbibliography,floatfix,superscriptaddress,showkeys]{revtex4-2} 
\usepackage{amsmath}
\usepackage{amsfonts}
\usepackage{amssymb}
\usepackage{hyperref}
\usepackage{tabulary}
\usepackage[dvipsnames]{xcolor}
\usepackage{soul}
\usepackage{siunitx} %
\usepackage{selinput}%
\usepackage{todonotes}
\usepackage{cleveref} %
\usepackage{upgreek} %

\usepackage{graphicx}
\usepackage{blindtext}
\usepackage{tabulary}
\usepackage{rotating}
\usepackage{afterpage}

\usepackage{multirow}
\usepackage{makecell}

\usepackage{booktabs}

\newcommand{\grey}[1]{\textcolor{gray}{#1}}

\begin{document}
\title{Tilt-to-length coupling in LISA Pathfinder: long-term stability}

\def\addressb{Max Planck Institute for Gravitational Physics (Albert-Einstein-Institut), 
30167 Hannover, Germany}
\def\addressluh{Leibniz Universit\"at Hannover, 30167 Hannover, Germany}
\def\addressbp{University of Potsdam, Institute of Physics and Astronomy, 14476 Potsdam, Germany}
\def\addressbn{Department of Physics, Aristotle University of Thessaloniki, Thessaloniki 54124, Greece}
\def\addressa{European Space Astronomy Centre, European Space Agency, Villanueva de la
    Ca\~{n}ada, 28692 Madrid, Spain}
\def\addressc{APC, Universit\'e de Paris, CNRS, Astroparticule et Cosmologie, F-75006 
    Paris, France}
\def\addresscy{IRFU, CEA, Universit\'e Paris-Saclay, F-91191 Gif-sur-Yvette, France}
\def\addressd{High Energy Physics Group, Physics Department, Imperial College London, 
    Blackett Laboratory, Prince Consort Road, London, SW7 2BW, UK }
\def\addresse{Dipartimento di Fisica, Universit\`a di Roma ``Tor Vergata'',  and INFN, 
    sezione Roma Tor Vergata, I-00133 Roma, Italy}
\def\addressf{Department of Industrial Engineering, University of Trento, via Sommarive 9, 38123 Trento, Italy
    and Trento Institute for Fundamental Physics and Application / INFN, 38123 Povo, Trento, Italy}
\def\addressh{European Space Technology Centre, European Space Agency, 
    Keplerlaan 1, 2200 AG Noordwijk, The Netherlands}
\def\addressi{Dipartimento di Fisica, Universit\`a di Trento and Trento Institute for 
    Fundamental Physics and Application / INFN, 38123 Povo, Trento, Italy}
\def\addressk{Istituto di Fotonica e Nanotecnologie, CNR-Fondazione Bruno Kessler, 
    I-38123 Povo, Trento, Italy}
\def\addressj{The School of Physics and Astronomy, University of
    Birmingham, Birmingham, UK}
\def\addressl{Institut f\"ur Geophysik, ETH Z\"urich, Sonneggstrasse 5, CH-8092, 
    Z\"urich, Switzerland}
\def\addressm{The UK Astronomy Technology Centre, Royal Observatory, Edinburgh, Blackford 
    Hill, Edinburgh, EH9 3HJ, UK}
\def\addressn{Institut de Ci\`encies de l'Espai (ICE, CSIC), Campus UAB, Carrer de Can 
    Magrans s/n, 08193 Cerdanyola del Vall\`es, Spain}
\def\addresso{DISPEA, Universit\`a di Urbino Carlo Bo, Via Santa Chiara, 27 61029 
    Urbino/INFN, Italy}
\def\addressp{European Space Operations Centre, European Space Agency, 64293 Darmstadt, 
    Germany}
\def\addressq{Physik Institut, 
    Universit\"at Z\"urich, Winterthurerstrasse 190, CH-8057 Z\"urich, Switzerland}
\def\addressr{SUPA, Institute for Gravitational Research, School of Physics and 
    Astronomy, University of Glasgow, Glasgow, G12 8QQ, UK}
\def\addresss{Department d'Enginyeria Electr\`onica, Universitat Polit\`ecnica de 
    Catalunya,  08034 Barcelona, Spain}
\def\addresst{Institut d'Estudis Espacials de Catalunya (IEEC), C/ Gran Capit\`a 2-4, 
    08034 Barcelona, Spain}
\def\addressu{Gravitational Astrophysics Lab, NASA Goddard Space Flight Center, 8800 
    Greenbelt Road, Greenbelt, MD 20771 USA}
\def\addressbb{Department of Mechanical and Aerospace Engineering, MAE-A, P.O. Box 
    116250, University of Florida, Gainesville, Florida 32611, USA}
\def\addressbbb{Department of Physics,
    2001 Museum Road, University of Florida, Gainesville, Florida 32611, USA}
\def\addressbh{Institut für Theoretische Physik, Universität Heidelberg, Philosophenweg 16, 69120 Heidelberg, Germany}
\def\addresscc{Istituto di Fotonica e Nanotecnologie, CNR-Fondazione Bruno Kessler, 
    I-38123 Povo, Trento, Italy}
\def\addressdd{isardSAT SL, Marie Curie 8-14, 08042 Barcelona, Catalonia, Spain}
\def\addressee{Escuela Superior de Ingenier\'ia, Universidad de C\'adiz, 11519 C\'adiz, 
    Spain}
\def\addressnn{Observatoire de la C\^{o}te d'Azur, Boulevard de l'Observatoire CS 34229 - 
    F 06304 NICE, France}
\def\addressff{City University of Applied Sciences, Flughafenallee 10, 28199 Bremen, 
    Germany}
\def\addressgg{Texas A\&M University, 701 H.R. Bright Bldg,
    College Station, TX 77843-3141 USA}
\def\addresshh{OHB System AG, Universit\"atsallee 27-29, 28359 Bremen, Germany}
\def\addressii{Airbus Defence and Space, Claude-Dornier-Strasse, 88090 Immenstaad, 
    Germany}
\def\addressrr{Department of Quantitative Methods, Universidad Loyola Andalucia, Avenida 
de las Universidades s/n, 41704, Dos Hermanas, Sevilla, Spain}

\author{M~Armano}\affiliation{\addressh}
\author{H~Audley}\affiliation{\addressb}\affiliation{\addressluh}
\author{J~Baird}\affiliation{\addressc}
\author{P~Binetruy}\thanks{Deceased 30 March 2017}\affiliation{\addresscy}%
\author{M~Born}\affiliation{\addressb}\affiliation{\addressluh}
\author{D~Bortoluzzi}\affiliation{\addressf}
\author{E~Castelli}\affiliation{\addressi}
\author{A~Cavalleri}\affiliation{\addresscc}
\author{A~Cesarini}\affiliation{\addresso}
\author{A\,M~Cruise}\affiliation{\addressj}
\author{K~Danzmann}\affiliation{\addressb}\affiliation{\addressluh}
\author{M~de Deus Silva}\affiliation{\addressa}
\author{I~Diepholz}\affiliation{\addressb}\affiliation{\addressluh}
\author{G~Dixon}\affiliation{\addressj}
\author{R~Dolesi}\affiliation{\addressi}
\author{L~Ferraioli}\affiliation{\addressl}
\author{V~Ferroni}\affiliation{\addressi}
\author{E\,D~Fitzsimons}\affiliation{\addressm}
\author{M~Freschi}\affiliation{\addressa}
\author{L~Gesa}\thanks{Deceased 29 May 2020}\affiliation{\addressn}\affiliation{\addresst}
\author{D~Giardini}\affiliation{\addressl}
\author{F~Gibert}\affiliation{\addressi}%
\author{R~Giusteri}\affiliation{\addressb}\affiliation{\addressluh}
\author{C~Grimani}\affiliation{\addresso}
\author{J~Grzymisch}\affiliation{\addressh}
\author{I~Harrison}\affiliation{\addressp}
\author{M-S~Hartig}\email{marie-sophie.hartig@aei.mpg.de}\affiliation{\addressb}\affiliation{\addressluh}
\author{G~Heinzel}\affiliation{\addressb}\affiliation{\addressluh}
\author{M~Hewitson}\affiliation{\addressb}\affiliation{\addressluh}
\author{D~Hollington}\affiliation{\addressd}
\author{D~Hoyland}\affiliation{\addressj}
\author{M~Hueller}\affiliation{\addressi}
\author{H~Inchausp\'e}\affiliation{\addressbh}%
\author{O~Jennrich}\affiliation{\addressh}
\author{P~Jetzer}\affiliation{\addressq}
\author{U~Johann}\affiliation{\addressii}
\author{B~Johlander}\affiliation{\addressh}
\author{N~Karnesis}\email{karnesis@auth.gr}\affiliation{\addressbn}
\author{B~Kaune}\affiliation{\addressb}\affiliation{\addressluh}
\author{C\,J~Killow}\affiliation{\addressr}
\author{N~Korsakova}\affiliation{\addressc}%
\author{J\,A~Lobo}\thanks{Deceased 30 September 2012}\affiliation{\addressn}\affiliation{\addresst}
\author{J\,P~L\'opez-Zaragoza}\affiliation{\addressn}
\author{R~Maarschalkerweerd}\affiliation{\addressp}
\author{D~Mance}\affiliation{\addressl}
\author{V~Mart\'{i}n}\affiliation{\addressn}\affiliation{\addresst}
\author{L~Martin-Polo}\affiliation{\addressa}
\author{F~Martin-Porqueras}\affiliation{\addressa}
\author{J~Martino}\affiliation{\addressc}
\author{P\,W~McNamara}\affiliation{\addressh}
\author{J~Mendes}\affiliation{\addressp}
\author{L~Mendes}\affiliation{\addressa}
\author{N~Meshksar}\affiliation{\addressl}
\author{M~Nofrarias}\affiliation{\addressn}\affiliation{\addresst}
\author{S~Paczkowski}\affiliation{\addressb}\affiliation{\addressluh}
\author{M~Perreur-Lloyd}\affiliation{\addressr}
\author{A~Petiteau}\affiliation{\addressc}\affiliation{\addresscy}
\author{E~Plagnol}\affiliation{\addressc}
\author{J~Ramos-Castro}\affiliation{\addresss}%
\author{J~Reiche}\affiliation{\addressb}\affiliation{\addressluh}
\author{F~Rivas}\affiliation{\addressrr}
\author{D\,I~Robertson}\email{david.robertson@glasgow.ac.uk}\affiliation{\addressr}
\author{G~Russano}\affiliation{\addressi}
\author{J~Sanjuan}\affiliation{\addressbbb}
\author{J~Slutsky}\affiliation{\addressu}
\author{C\,F~Sopuerta}\affiliation{\addressn}\affiliation{\addresst}
\author{T~Sumner}\affiliation{\addressd}\affiliation{\addressbbb}
\author{D~Texier}\affiliation{\addressa}
\author{J\,I~Thorpe}\affiliation{\addressu}
\author{D~Vetrugno}\affiliation{\addressi}
\author{S~Vitale}\affiliation{\addressi}
\author{G~Wanner}\email{gudrun.wanner@aei.mpg.de}\affiliation{\addressluh}\affiliation{\addressb}
\author{H~Ward}\affiliation{\addressr}
\author{P\,J~Wass}\affiliation{\addressd}\affiliation{\addressbb}
\author{W\,J~Weber}\affiliation{\addressi}
\author{L~Wissel}\affiliation{\addressb}\affiliation{\addressluh}
\author{A~Wittchen}\affiliation{\addressb}\affiliation{\addressluh}
\author{P~Zweifel}\affiliation{\addressl}

\date{\today}

\begin{abstract}
The tilt-to-length coupling during the LISA Pathfinder mission has been numerically and analytically modeled for particular timespans. In this work, we investigate the long-term stability of the coupling coefficients of this noise.
We show that they drifted slowly (by 1\,$\upmu$m/rad and 6$\times10^{-6}$ in 100 days) and strongly correlated to temperature changes within the satellite (8\,$\upmu$m/rad/K and 30$\times10^{-6}$/K). 
Based on analytical TTL coupling models, we attribute the temperature-driven coupling changes to rotations of the test masses and small distortions in the optical setup.
Particularly, we show that LISA Pathfinder's optical baseplate was bent during the cooldown experiment, which started in late 2016 and lasted several months.
\end{abstract}
\maketitle

\section{Introduction}
\label{sec:Intro}

In July 2017, the LISA Pathfinder (LPF) mission ended \cite{McNamara2008,Armano2016,Armano2018}.
During the previous 19 months, LPF had successfully demonstrated the necessary dynamical stability of its hosted free-floating test masses and relevant technologies for the Laser Interferometer Space Antenna (LISA), the first space-based gravitational wave observatory \cite{Danzmann2011,LISAmission,elisa13ARXIV}.
LPF exceeded its requirements by several orders of magnitude \cite{Armano2016,Armano2018}.

One of the main noise sources investigated during the LPF mission was the optical cross-talk of angular and translational spacecraft (S/C) jitter into the interferometric length signal. This noise is called tilt-to-length (TTL) coupling and is also expected to be a major noise source in LISA \cite{Wanner2024,Paczkowski2022,George2022,Wegener2024,Houba2022a,Houba2022b}.
During the LPF mission, it was successfully reduced by realignment. The residual coupling was then subtracted in post-processing \cite{Wanner2017}. 
It was further shown in \cite{LPFana22} that the TTL coupling in LPF can be modeled analytically.
When applying this model to the LPF data, it describes well how the level of TTL coupling changed in response to test mass realignments. However, the initial magnitude of the coupling depended on partly unknown parameters defined by the setup and nominal test mass alignment \cite{LPFdata22}.
These parameters changed during the time of the mission and, therefore, affected the TTL coupling on a long-term scale. 

In this paper, we investigate the computed coupling coefficients' stability for the LPF mission's time.
We show that they were not entirely stable but drifted slowly and showed a strong dependency on temperature changes onboard. While the absolute coefficient changes were smaller than predicted for LISA, we find significant relative alterations. 
Moreover, we discuss the implications of our analysis for the stability of the optical setup and what we learn from it for the upcoming LISA mission. 

We start in Sec.~\ref{sec:Basics} with a short explanation of the TTL coupling sources and mechanisms in LPF (Sec.~\ref{sec:TTLinLPF}). 
The corresponding formulas modeling this TTL coupling are then introduced in Sec.~\ref{sec:TTLmodels}. We differentiate between a model gained by a data fit algorithm and an analytically derived one.
Furthermore, we briefly explain the different jitter measurement techniques and control schemes in LISA Pathfinder (\ref{sec:DWS+DFACS}).
In Sec.~\ref{sec:CoeffStability}, we present our data analysis of the long-term behavior of the fitted coupling coefficients. We show how these were correlated to changes in the temperature inside the S/C, the angular readout and the spot positions on the detectors.
Combining these observations with our knowledge about the coupling coefficients from the analytical TTL coupling model, we discuss their implications for the optical setup stability in Sec.~\ref{sec:OBstability}.
By this, we can show that the optical bench (OB) was subject to distortion during a cooldown experiment in the second half of the mission.
Finally, our results are summarized in Sec.~\ref{sec:summary}.

\section{Basics}
\label{sec:Basics}

\subsection{Tilt-To-Length Coupling in LISA Pathfinder}
\label{sec:TTLinLPF}

\begin{figure}
\centering
\includegraphics[width=0.86\columnwidth]{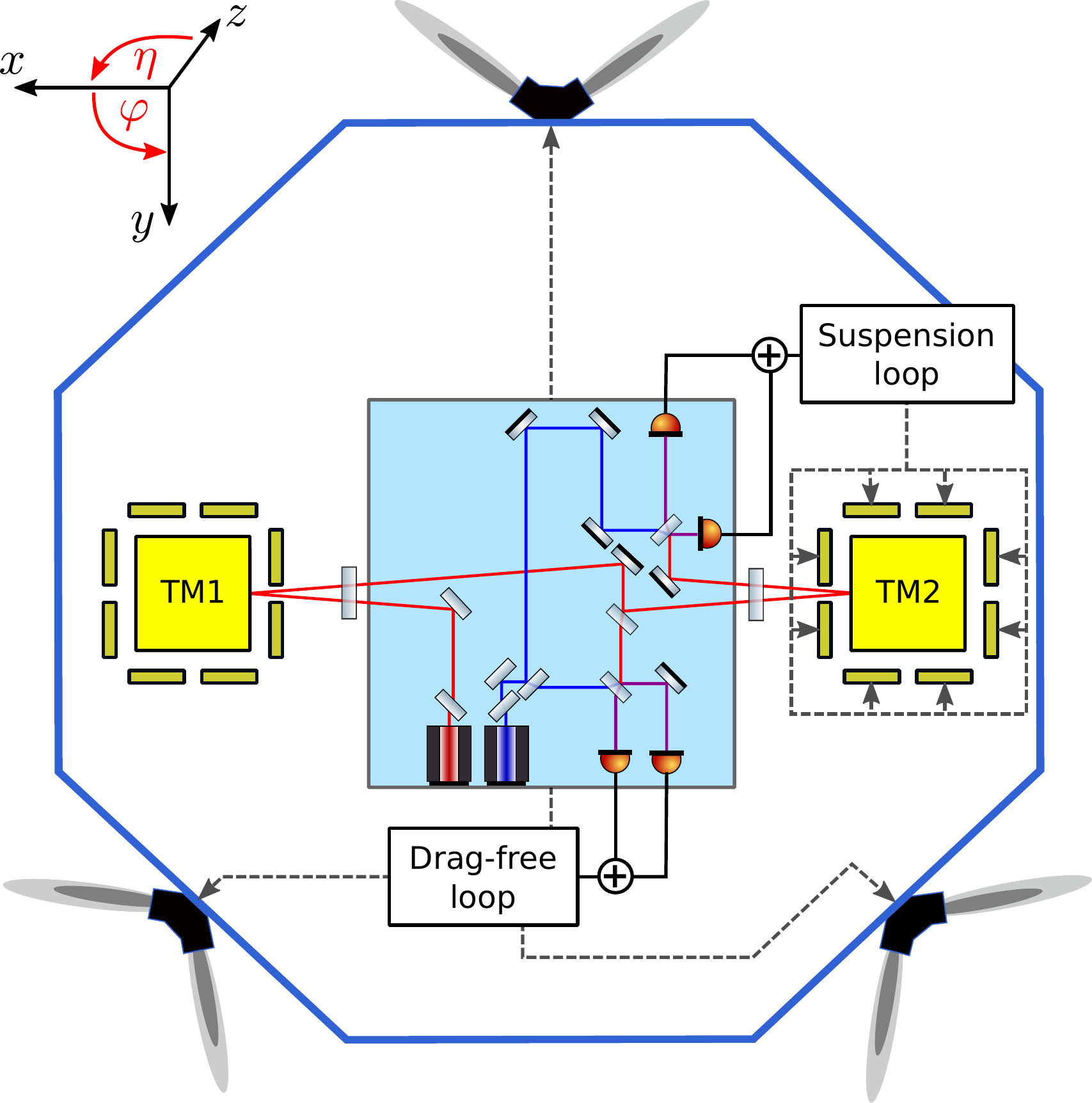}
  \caption{Sketch of the interferometers onboard LPF used to measure the relative positions of the S/C and the TMs: 
  The two diodes shown at the bottom edge of the OB (light blue) formed the x1-interferometer, which measured the relative alignment of the first TM (TM1) and the S/C. The drag-free control loop utilized this measurement to preserve the alignment of the S/C relative to TM1 via the thrusters. 
  The two other diodes built the x12-interferometer, which yielded the relative displacement of the TMs and, additionally, the second TM (TM2) alignment with respect to the S/C. This measurement was fed into the suspension control loop, which made the TM2 follow the S/C via the electrodes inside the EH. 
  This figure is a reprint from \cite{LPFdata22}.}
\label{fig:LPFscetch}
\end{figure}

The main scientific setup in LPF consisted of two test masses (TMs) which were freely floating inside an electrode housing (EH) each. 
Between the two EHs, an OB with the optical setup was placed, see Fig.~\ref{fig:LPFscetch}. 
The single-link measurement in LISA was simulated by this setup, with one of the TMs replacing the TM in the far S/C. 
The primary signal to be measured was the change in distance between (or relative acceleration of) the two TMs. This measurement was performed interferometrically. 
Therefore, one laser beam (red in Fig.~\ref{fig:LPFscetch}) was reflected at both TMs and then interfered with a second beam (blue in Fig.~\ref{fig:LPFscetch}) that stayed on the optical bench and remained unchanged under S/C or TM movements.
 
The jitter of either the S/C or the TMs changed the direction and optical path length of the beam that was reflected by the TMs.
Correspondingly, the interference signal changed, although the distance between both TMs had not changed.
We refer to this cross-talk noise as TTL coupling.
In LPF, it was mostly driven by S/C jitter and was the dominating noise source between 20 and 200\,mHz before its subtraction in post-processing \cite{Armano2016,LPFdata22}.

The most significant TTL coupling contributors were the translational S/C jitters orthogonal to the optical axis from one TM to the other one (``lateral" jitter in $y$- and $z$-direction) and the angular S/C jitter in yaw ($\varphi$) and pitch ($\eta$) \cite{LPFdata22}. The corresponding coordinate system is shown in Fig.~\ref{fig:LPFscetch}.
Lateral S/C jitter caused a beam walk along the TM surfaces, which changed - for a nominally tilted TM - the beam's optical pathlength \cite{G22,LPFana22}.
The same happened, to a slightly smaller degree, for angular S/C jitter. In this case, the beam accumulated an additional optical pathlength change due to its changes in direction after the reflection at the TMs. Also, the point of incidence at the detector significantly shifted, altering the differential wavefront of the interfering beams and, therefore, the signal \cite{G22,NG22,LPFana22}.

In the following section, we present how we modeled the TTL coupling in LPF.

\subsection{The TTL models}
\label{sec:TTLmodels}

In the analysis presented in this work, we make use of two TTL coupling models: 
the fit model that has been used during the LPF mission for the subtraction of the TTL noise in post-processing \cite{Wanner2017} 
and a model derived analytically, that provides the dependency of the coupling on the alignment of the TMs \cite{LPFana22}.
Both models describe how the S/C jitter in the different degrees of freedom (DoF) coupled into the measurement of the relative TM accelerations along the optical axis $\Delta g_\text{xacc}$.

For the fit of the TTL coupling contributions, the coupling of the accelerations in the four orthogonal DoF ($\varphi,\,\eta,\,y,\,z$) is being considered.
The corresponding motion of the S/C was approximated from the x1- and x12-interferometer (Fig.~\ref{fig:LPFscetch}) measurements as well as the electro-statical readouts (see Sec.~\ref{sec:DWS+DFACS}):
$j_\text{SC}\approx\bar{j}=(j_1+j_2)/2$, $j\in\{\varphi,\eta,y,z\}$. 
More precisely, these variables describe the motion of the SC relative to the TMs. However, during the timespans used for the presented analyses, the TMs were in free-fall with residual accelerations in the \SI{}{fm/s^2} regime. Therefore, the observed acceleration originates predominantly from SC motion.
In addition, the fit also takes into account the accelerations due to lateral displacements of the TMs via stiffness and the residual jitter of the S/C along the sensitive $x$-axis ($\ddot{o}_1$), which did not change the distance between the TMs but coupled into the $\Delta g_\text{xacc}$ signal due to imperfections in the setup symmetry.
The resulting model is then 
\begin{align}
\begin{split}
\Delta g_\text{xacc}^\text{fit} = &\
C_\varphi^\text{fit}\,\ddot{\overline{\varphi}} 
+ C_\eta^\text{fit}\,\ddot{\overline{\eta}} 
+ C_y^\text{fit}\,\ddot{\overline{y}} 
+ C_z^\text{fit}\,\ddot{\overline{z}} \\
&+ C_{y,s}^\text{fit}\,\overline{y} 
+ C_{z,s}^\text{fit}\,\overline{z}
+ C_{o_1}^\text{fit}\,\ddot{o}_1 \,,
\end{split}
\label{eq:fitmodel}
\end{align}
The TTL coupling coefficients $C^\text{fit}$ are derived by fitting the shown model to mission data.
The first four terms were dominant in all time segments analyzed by us. 

The analytically derived model only considers these dominant terms, i.e.
\begin{align}
\Delta g_\text{xacc}^\text{ana} = 
C_\varphi^\text{ana}\,\ddot{\overline{\varphi}} 
+ C_\eta^\text{ana}\,\ddot{\overline{\eta}} 
+ C_y^\text{ana}\,\ddot{\overline{y}} 
+ C_z^\text{ana}\,\ddot{\overline{z}} \,.
\label{eq:anamodel}
\end{align}
The coupling coefficients $C^\text{ana}$ have been derived analytically \cite{LPFana22}. 
For their computation, all geometric and non-geometric (wavefront and detector geometry related) TTL coupling mechanisms adding up in the case of S/C jitter in LPF have been taken into account.
This yielded 
\begin{subequations}
\begin{align}
C_\varphi^\text{ana} &= C_{\varphi,0}
+0.210^{+0.017}_{-0.016}\,\frac{\mathrm{m}}{\mathrm{rad}^2} \,\varphi_{1}
+0.182^{+0.018}_{-0.020}\,\frac{\mathrm{m}}{\mathrm{rad}^2} \,\varphi_{2} \label{eq:anaCphi}\\
C_\eta^\text{ana} &= C_{\eta,0}
+0.209^{+0.017}_{-0.015}\,\frac{\mathrm{m}}{\mathrm{rad}^2} \,\eta_{1}
 +0.178^{+0.018}_{-0.019}\,\frac{\mathrm{m}}{\mathrm{rad}^2}\,\eta_{2} \label{eq:anaCeta}\\
C_y^\text{ana} &= C_{y,0}
+\, 1.000^{+0}_{-0} \ \frac{1}{\text{rad}}\,(-\varphi_{1}+\varphi_{2}) \label{eq:anaCy}\\
C_z^\text{ana} &= C_{z,0}
+\, 1.000^{+0}_{-0} \ \frac{1}{\text{rad}}\,(\eta_{1}-\eta_{2}) \,, \label{eq:anaCz}
\end{align}
\label{eq:anaC}%
\end{subequations}%
where the constant offsets $C_{j,0},\,j\in\{\varphi,\eta,y,z\},$ depended on the setup and nominal TM alignment parameters, which are partly unknown and could change during the mission.
The angular readouts $\varphi_1,\,\varphi_2,\,\eta_1,\,\eta_2$ denote the (re-)alignments of the two TMs. 
The TM alignment readouts available on LPF are introduced in the following subsection. 

It has been shown in \cite{LPFdata22} that both models successfully describe the coupling of TTL noise into $\Delta g_\text{acc}$. 
Thus, we make use of both in the following analyses.

\subsection{The jitter readout and control}
\label{sec:DWS+DFACS}
For the description of the long-term stability of the TTL coupling, we need to know how the jitter and the TM alignments were measured on LPF. Also, the satellite's attitude control system, which was using these measurements is of particular interest. We make use of our understanding of these systematics to explain the TTL coupling changes in the main part of this paper.

\subsubsection{Angular readout}
In LPF, three different angular readouts were available:
\begin{enumerate}
\item Differential wavefront sensing (DWS) \cite{Wanner2012,Morrison1994}: \\
  The interfering measurement (beam reflected at the TMs) and reference beam were detected by the x1- and x12-interferometers.
  The relative angle between these beams was computed from the difference of the phases measured by the four photodiode quadrants.
  The DWS signal provided the most precise angular readout among the three investigated ones. 
  We used these readouts for estimating the SC tilts in the TTL coupling models (Eq.~\eqref{eq:fitmodel} and Eq.~\eqref{eq:anamodel}) and the TM alignments in our analytical analysis (Eqs.~\eqref{eq:anaC}).
\item The gravitational reference sensor (GRS) \cite{Dolesi2003,Armano2020_actua-electr}: \\
  The angular alignment of the TMs relative to their housings was measured electrostatically. 
  Electrodes, which were installed at the inside of the EHs, measured variations in the distance between the housing wall and the test mass surfaces.
\item Differential power sensing (DPS) \cite{Wanner2012}: \\
  The relative beam alignment was computed via the power distribution over the four quadrants of the photodiodes. 
  An increasing power at one side of the quadrant photodiode was interpreted as a beam walk. By geometric dependencies, this beam walk can be related to test mass tilt: As the TM tilts, the reflected beam records this angle. For small angles, the beam walk would equal the beam's propagation distance between the TM and the detector times the angular change of the beam.
  Like the DWS, the DPS used the readouts of the x1- and x12-interferometers.
\end{enumerate}

\subsubsection{Lateral readout}
For the lateral jitter of the TMs only the GRS readout was available.
Like for the angular readout, the electrodes inside the housings were used for this measurement.
We used these readouts for the approximation of the lateral SC motion in Eq.~\eqref{eq:fitmodel} and Eq.~\eqref{eq:anamodel}.

\subsubsection{Control loops}
The alignment of the optical system was controlled by the drag-free attitude control system (DFACS). It locked the alignment of the satellite to predefined set-points of the much quieter TMs and corrected differential movements of the TMs. 
This control scheme utilized the lateral GRS-signals and either of the angular readouts.
The DWS angles were used for most measurement runs. 
The DFACS preserved the overall alignment and stability of the setup with two control loops, which are sketched in Fig.~\ref{fig:LPFscetch}.
Nominally, the drag-free control loop made the SC follow TM1 and the suspension loop controlled the TM alignment with respect to the SC.

\section{Long-Term Stability of the TTL Coupling Coefficients}
\label{sec:CoeffStability}

\begin{figure*}
\flushright
  \includegraphics[scale=0.295]{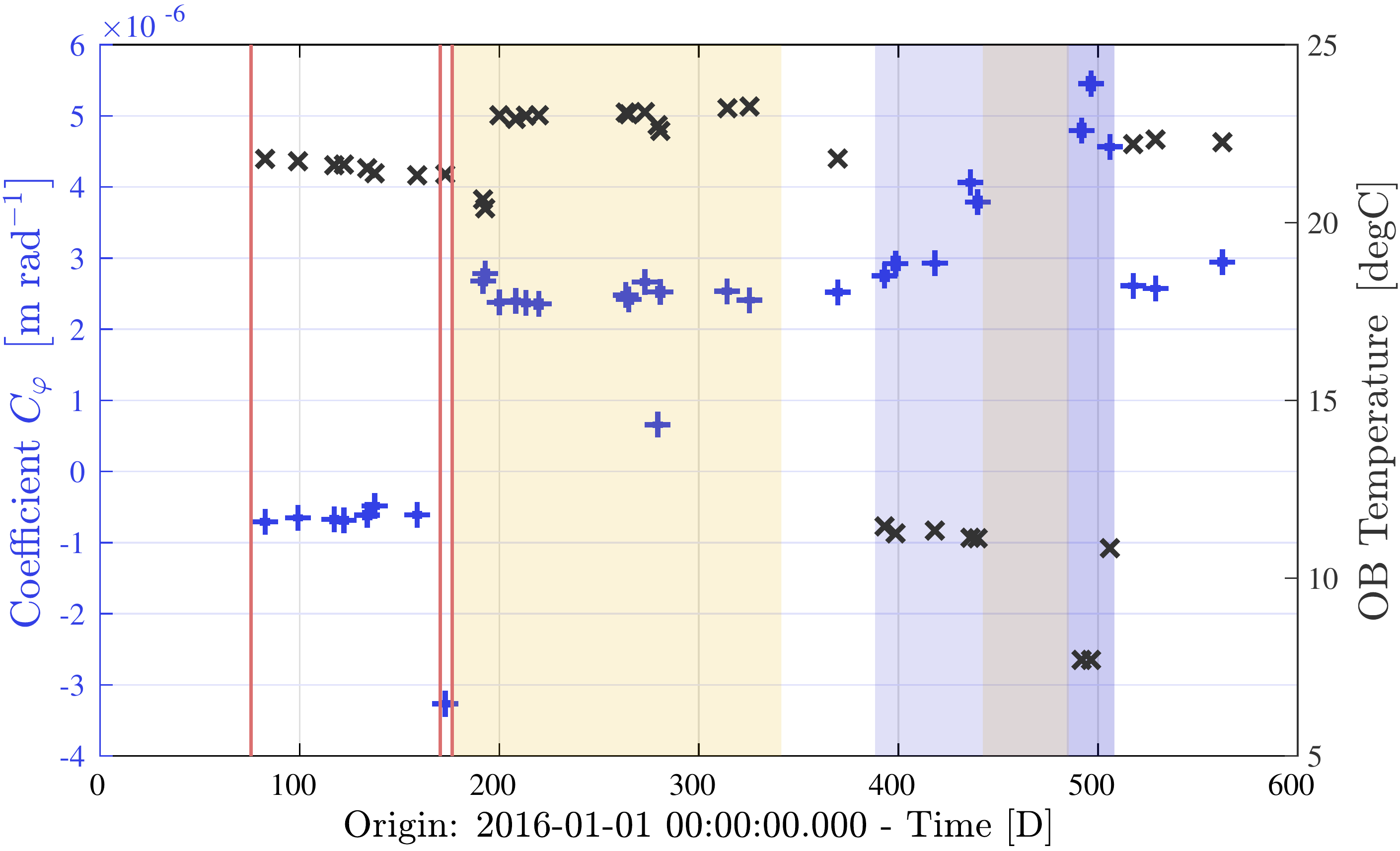} \quad %
  \includegraphics[scale=0.295]{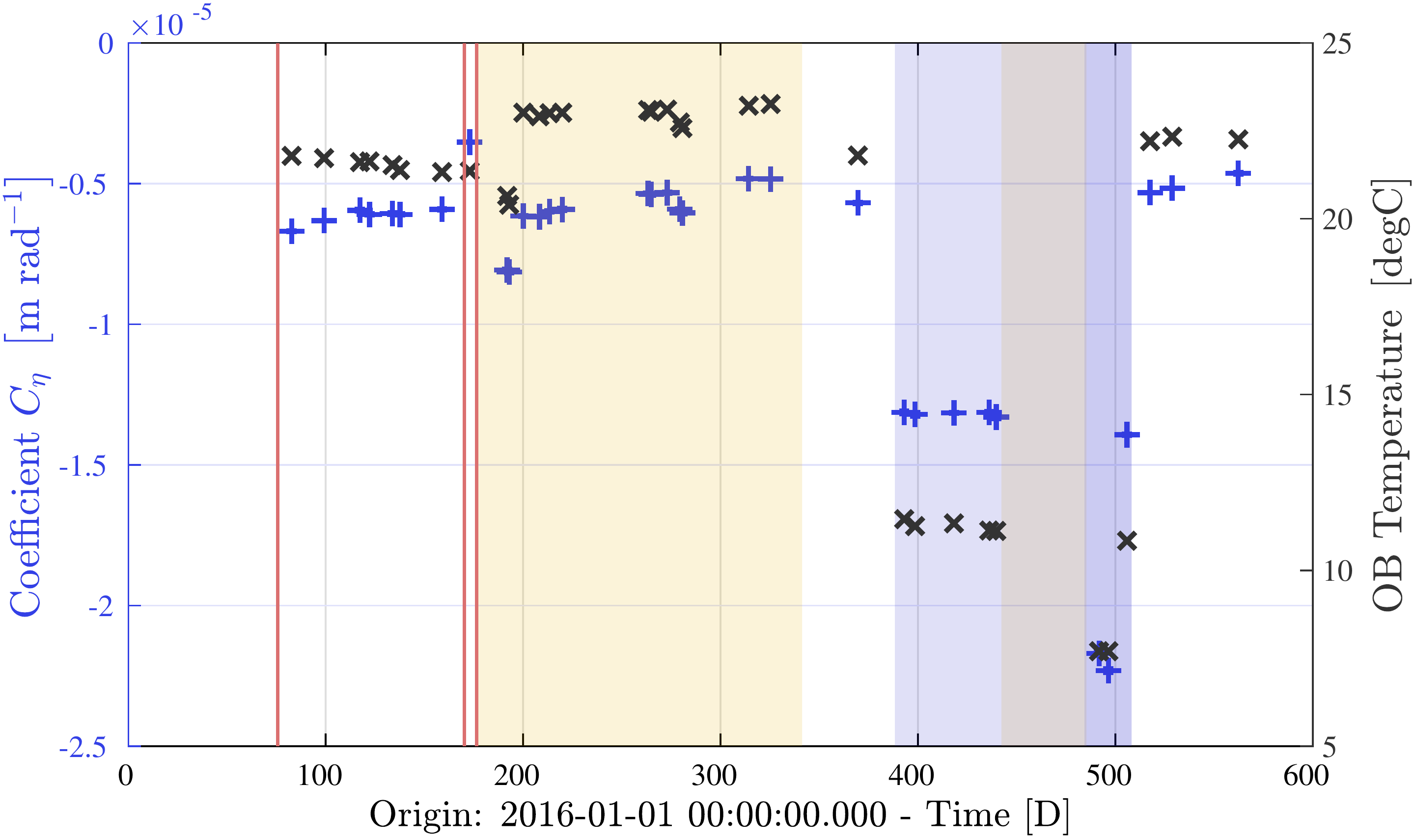} \\[0.5ex]
  \includegraphics[scale=0.295]{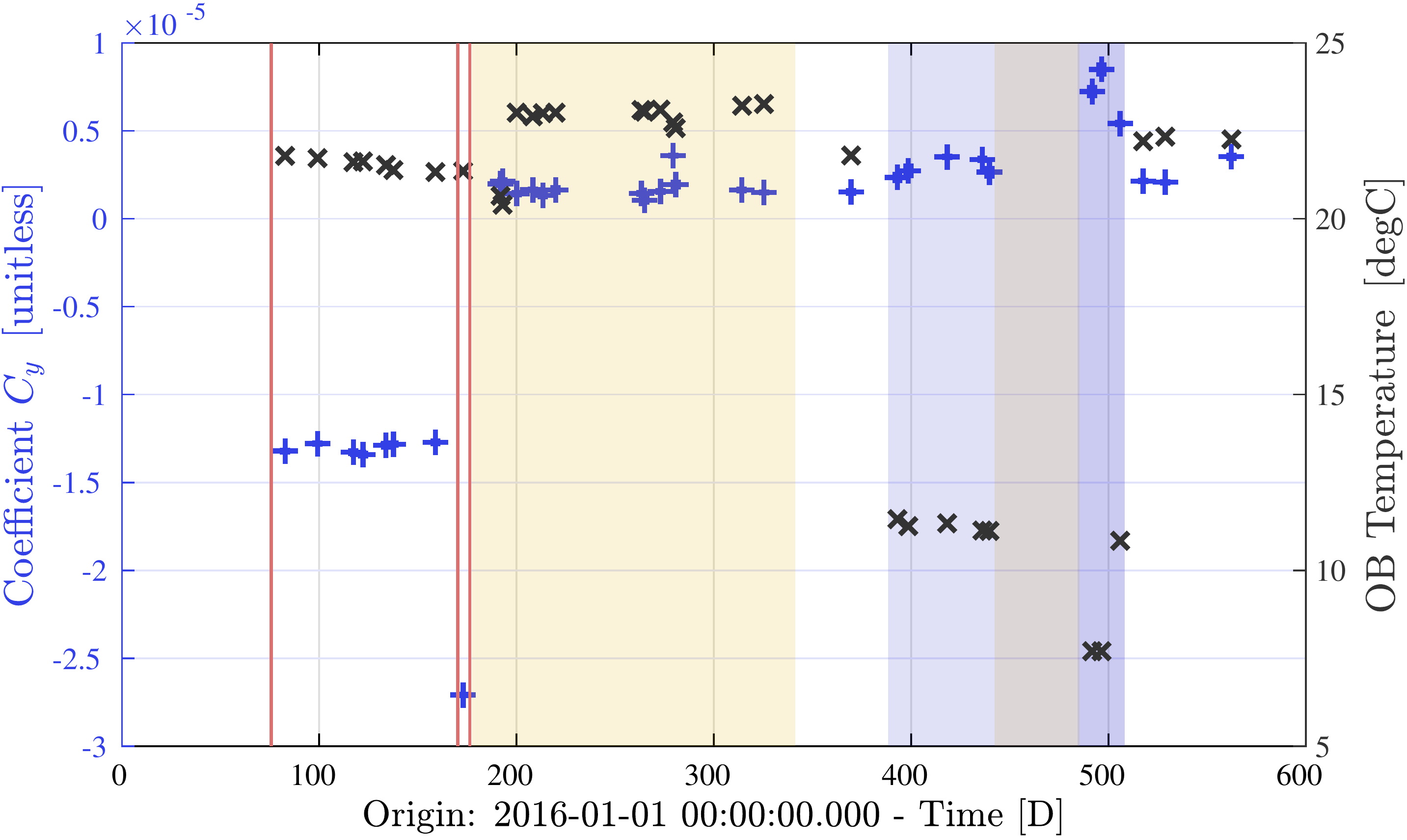} \quad\ \ %
  \includegraphics[scale=0.295]{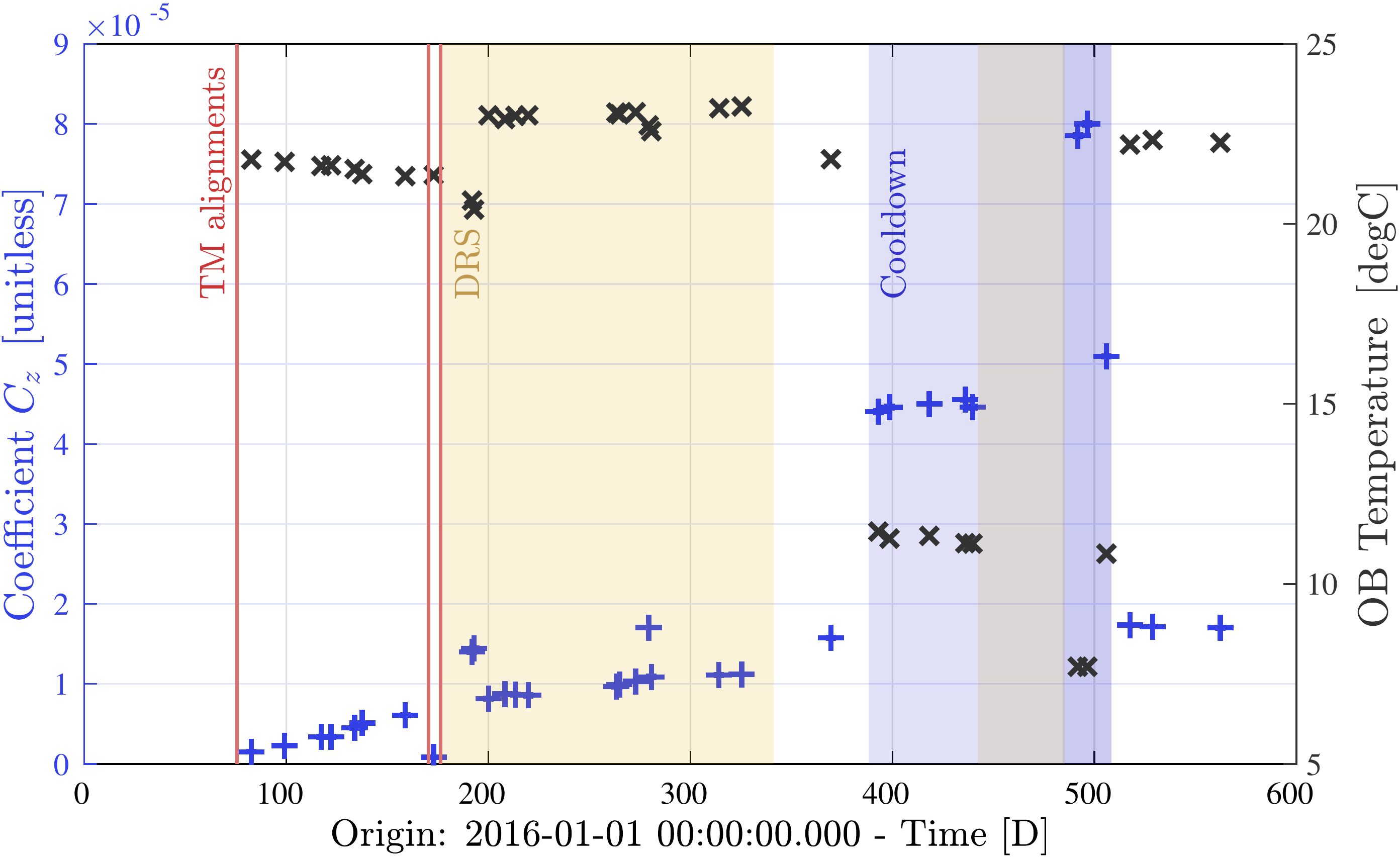}
  \caption{Long-term behavior of the fitted TTL coupling coefficients (blue pluses) and the mean OB temperature for the same timespans (black crosses, see Tab.~\ref{tab:timespans}). 
  The red lines mark the times of TM realignment for TTL minimization. The orange areas identify the times of the disturbance reduction system (DRS) \cite{DRS2018} operations, while during all other times, the LISA Technology Package (LTP) \cite{Armano2015} was used. The blue areas show the times of the cooldown experiment. The second DRS phase took place during the cooldown. After this DRS operation, the temperature was further decreased in LPF, reaching temperatures below the scale of the sensors. The actual temperatures reached approximately \SI{2}{\degree C}.}
\label{fig:TTLcoeffs}
\end{figure*}

For the investigation of the long-term stability of the TTL coupling coefficients, we fitted the coupling coefficients for the timespans with ultra reduced low authority (URLA) \cite{Giusteri2017,Armano2019_actuation}, and a setup alignment control utilizing the DWS angles.
The related timespans are summarized in Tab.~\ref{tab:timespans}. 
The fitted coupling coefficients for the lateral and angular acceleration are plotted in Fig.~\ref{fig:TTLcoeffs}.
The remaining three coefficients for stiffness ($C_y,\,C_z$) and SC-longitudinal motion ($C_{o_1}$) contributed only secondarily and are presented for completeness in the appendix (App.~\ref{sec:CoeffStability_CoeffsApp}).

In the following, we characterize the stability of the coupling coefficients and their temperature-dependency (Sec.~\ref{sec:CoeffStability_coeffs}).
Then, we discuss their relation to the readouts of the angular TM alignments (Sec.~\ref{sec:CoeffStability_angles}) and the beams' spot positions on the detector (Sec.~\ref{sec:CoeffStability_spot}).
Our observations are summarized and interpreted based on the analytical TTL model in Sec.~\ref{sec:CoeffStability_interpretation}.

\subsection{Characterisation of the Coefficient Stability}
\label{sec:CoeffStability_coeffs}

In Fig.~\ref{fig:TTLcoeffs}, we see significant variations of the coupling coefficients (blue pluses, left $y$-axis) throughout the mission time.
First, the coefficients changed in response to the TM realignments (red lines), which was expected and the aim of the realignments. For further details, see \cite{LPFdata22}.
In between and after these realignments, the coefficients drifted.
The lateral coefficients, i.e.\ $C_y$ and $C_z$, drifted by less than 6$\times10^{-6}$ and the angular coefficients, i.e.\ $C_\varphi$ and $C_\eta$, by less than 1\,$\upmu$m/rad in 100 days. 
Furthermore, we see significant changes (`jumps') at the beginning and end of the cooldown (blue areas).
Both observations were stronger for the coefficients scaling the jitter in the $xz$-plane, i.e.\ $C_\eta$ and $C_z$, than for the two coefficients in the orthogonal $xy$-plane.

When comparing the coefficients with the mean temperature measurements on the OB (black crosses, right $y$-axis), we find that the coefficient changes are related to the temperature changes onboard LPF.
This relation is most evident in the case of the jumps due to the beginning of the cooldown of the S/C.
At the end of January 2016, the temperature inside LPF was decreased by about \SI{10}{\degree}. After the second DRS phase, which took place during the cooldown,
a further decrease followed. During this time, the temperatures exceeded the measurement regime of the temperature sensors used for Fig.~\ref{fig:TTLcoeffs}.
Extrapolations suggest a decrease by not \SI{2}{\degree C} but \SI{10}{\degree C} \cite{Armano2019_temperature}. 
The coupling coefficients $C_\eta$ and $C_z$ significantly jumped when the satellite was cooled down. The coefficient change is approximately the same for both temperature decreases indicating a linear relationship.
After the cooldown, the coupling in these DoFs was again at the same level as before this experiment. 
Changes in the $C_\varphi$ and the $C_y$ coefficient were less significant.

A similar response of the coupling coefficients to the temperature changes can be observed between the third TM realignment and the cooldown.
In general, the temperature was about \SI{2}{\degree C} higher during the DRS phase compared to the LTP operations.
However, the temperature temporarily dropped during the measurement runs assigned to the 9th, 10th, 18th and 19th data points (days 191, 192, 278 and 280 in Fig.~\ref{fig:TTLcoeffs}). Simultaneously, the coupling coefficients differed from the mean level of the other coupling coefficients during the respective operation phases.
Also, after the DRS phase but before the cooldown, when the temperature was decreased to its original temperature level, the coefficients jumped again.
The sign of most coefficient changes was the same as for the jumps due to the cooldown. The only exception is the $C_\varphi$ coefficient for the measurement run starting at day 278, which we cannot explain.

Besides the coefficient jumps, we see a drift of the $C_z$ coefficient between the TM alignments and between the alignments and the cooldown. 
During this time, the temperature was stable except of short term shifts, and hence cannot explain the coefficient drift. 
A possible explanation could be a long-term mechanical stress relief concerning the optical setup.

In general, stresses and relaxations can cause distortions in optical systems.
These can be long-term effects or the result of environmental changes, e.g., alterations in the particle density, electro-statical charging, temperature.
We assume such relaxations caused the coefficient changes shown in Fig.~\ref{fig:TTLcoeffs} and investigate their origins further in the following sections.

\subsection{Discussion of the Angular Readouts}
\label{sec:CoeffStability_angles}

\begin{figure}
\flushright
  \includegraphics[scale=0.3]{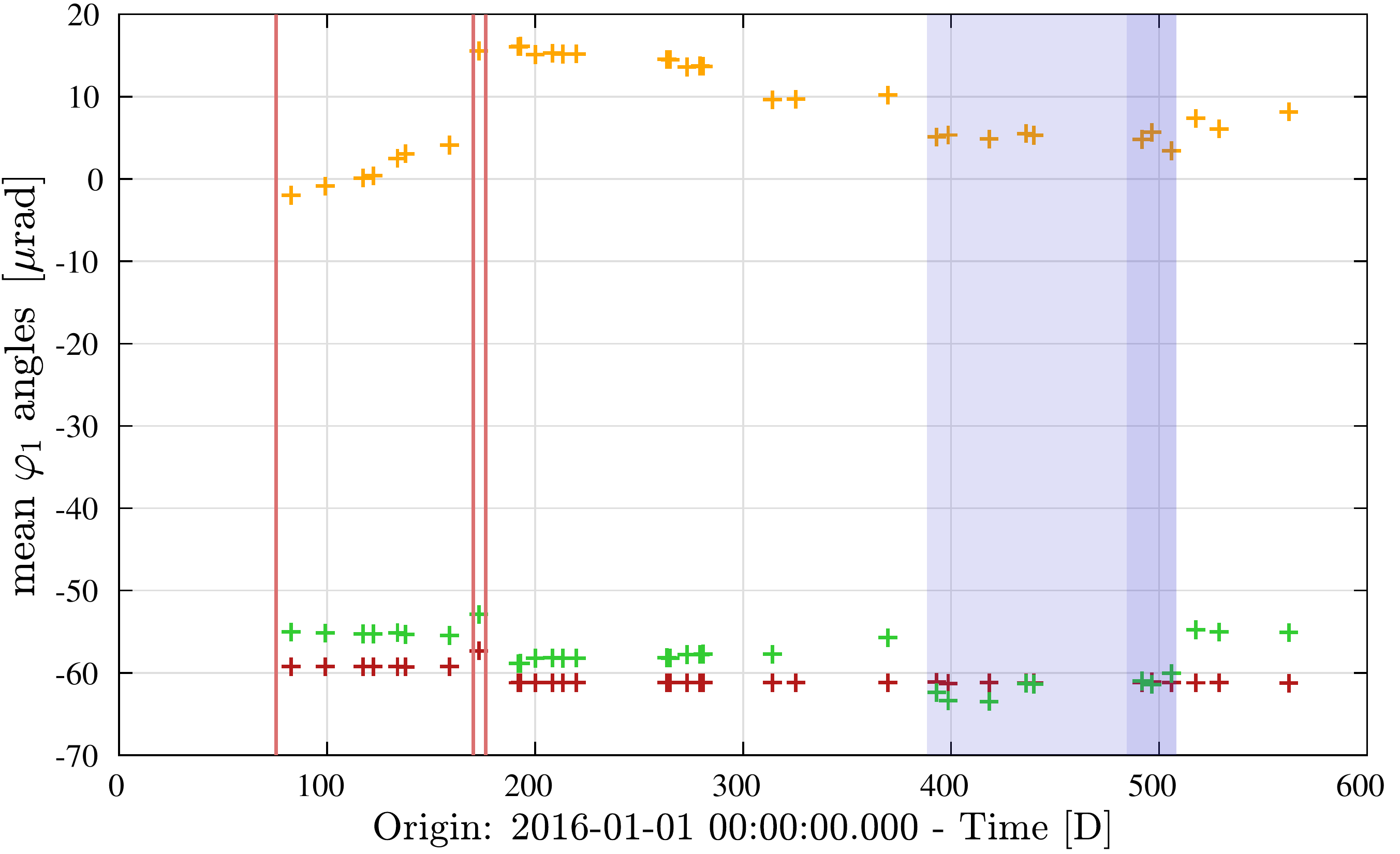} \\[1ex]
  \includegraphics[scale=0.3]{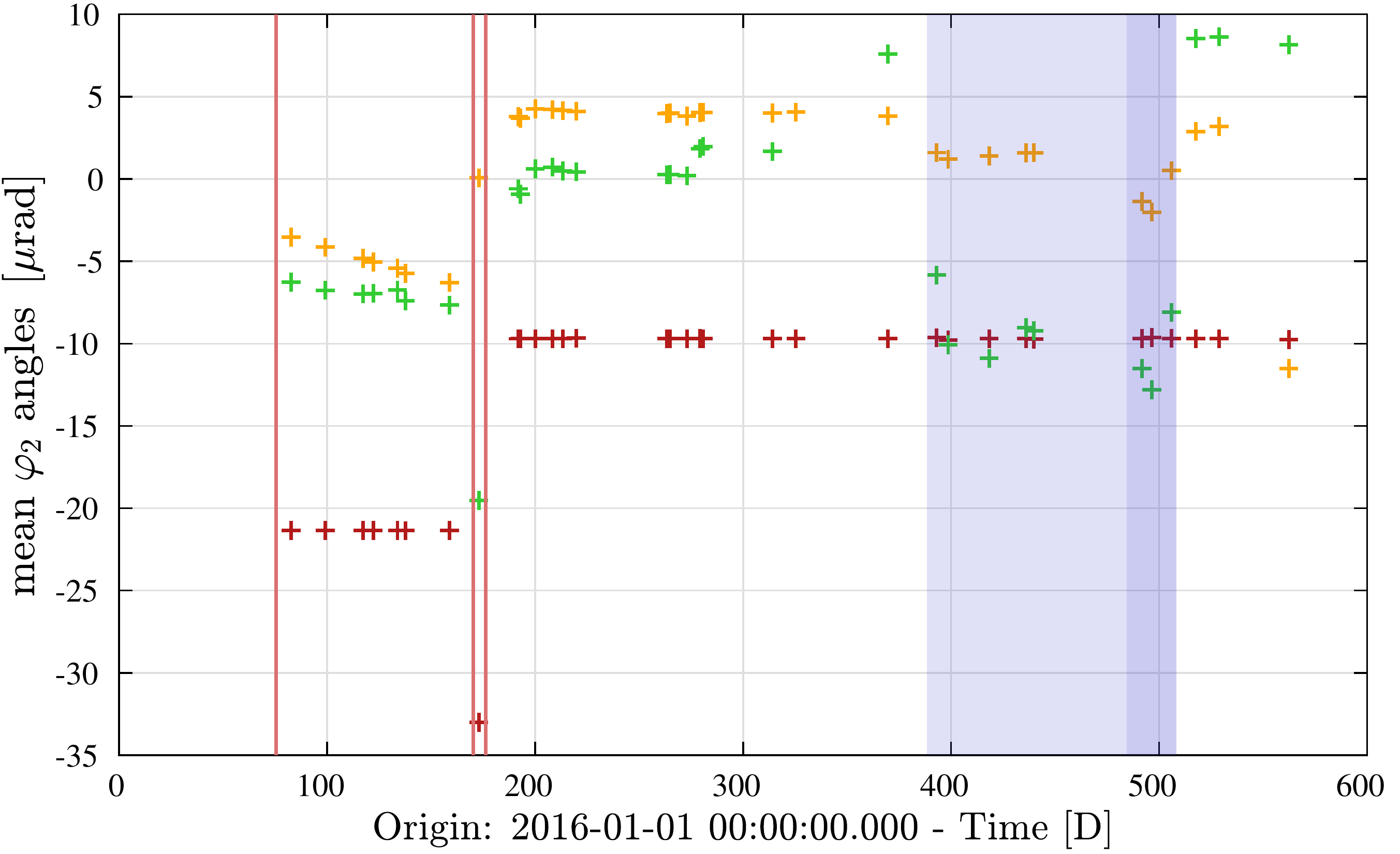} \\[1ex]
  \includegraphics[scale=0.3]{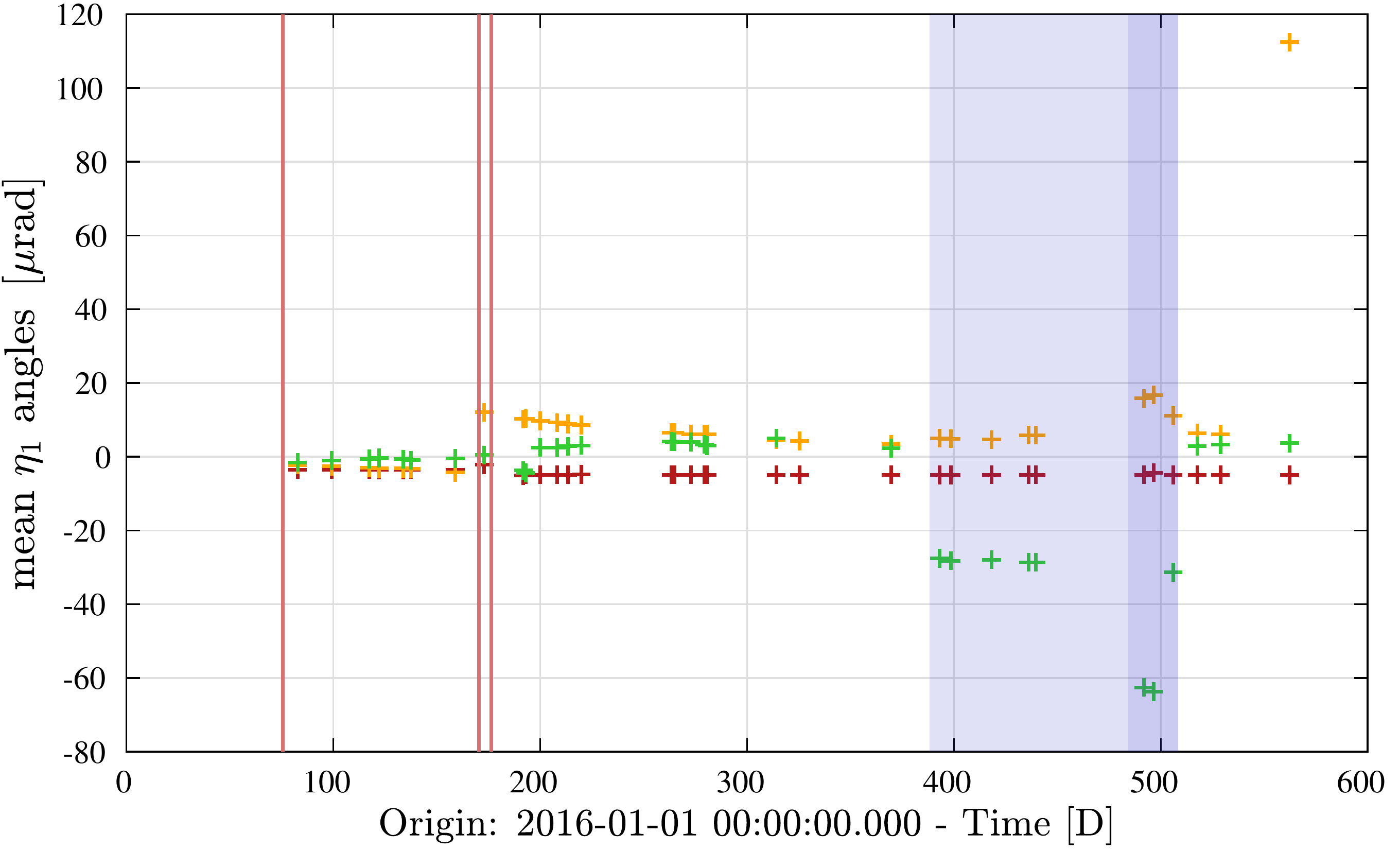} \\[1ex]
  \includegraphics[scale=0.3]{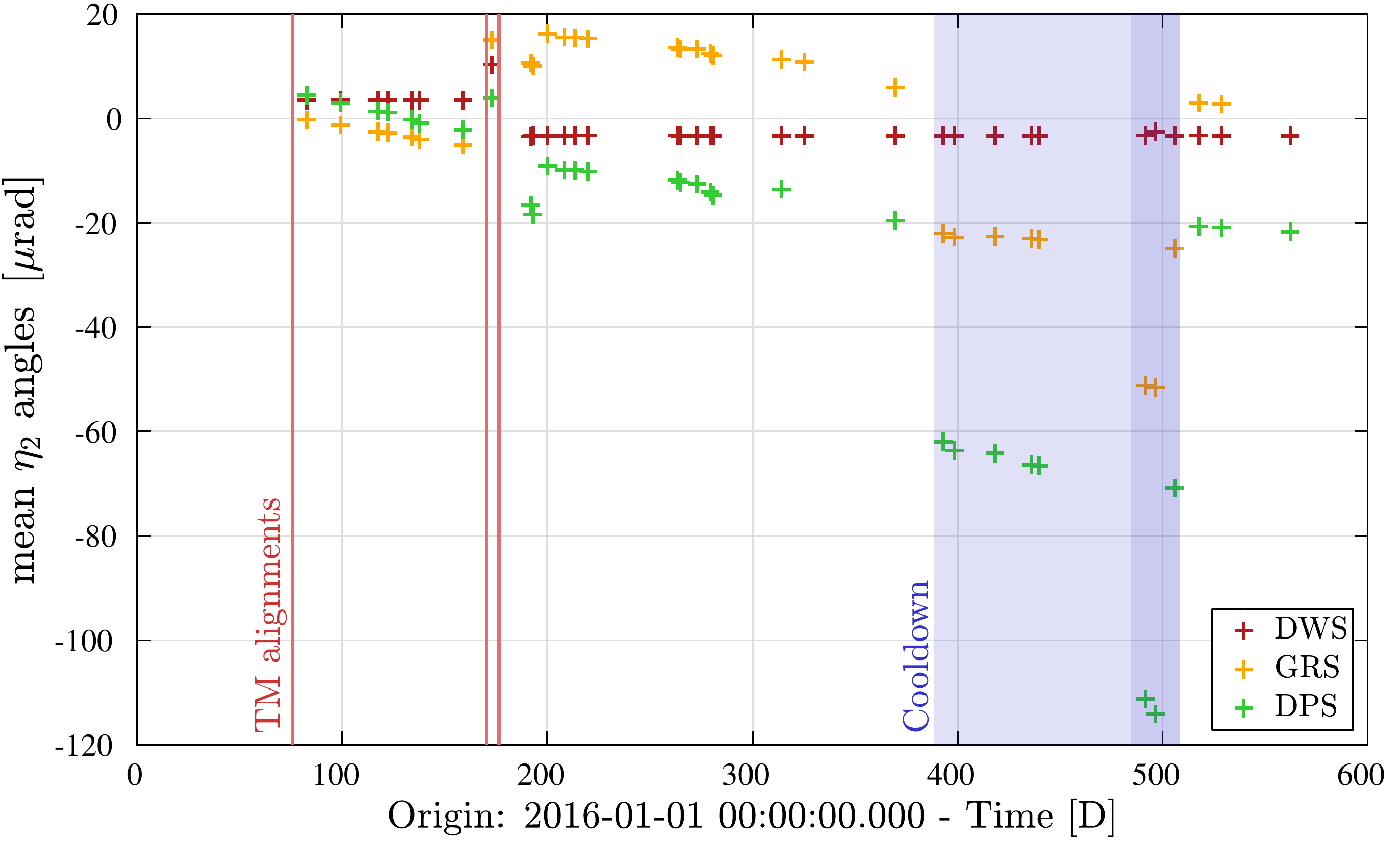}
  \caption{Mean angular readouts during the noise runs, which were considered for the computation of the TTL coupling coefficients in Fig.~\ref{fig:TTLcoeffs}. 
  The red pluses show the DWS readout which has been used to control the alignment stability of the setup. The GRS (yellow) and DPS (green) readouts are less stable. 
  The last GRS angle deviates for unknown reasons.}
\label{fig:TTLangles}
\end{figure}

To understand and interpret the long-term changes of the TTL coupling coefficients, we show the mean angular readouts at the investigated times (Tab.~\ref{tab:timespans}) in Fig.~\ref{fig:TTLangles}. 
In general, we find deviations between the three angular readouts.
In all of these measurement runs, the DWS angles were used for the system alignment control, i.e.\ the S/C and TM behavior was locked to these set-points. Correspondingly, 
the DWS angles only changed at the times of the TM alignment for TTL minimization (red lines) and remained constant afterwards. %
In contrast, the GRS and the DPS angles drifted in the long-term and jumped due to the cooldown of the satellite.
These effects were more significant for the pitch angles ($\eta_1,\,\eta_2$) than for the yaw angles ($\varphi_1,\,\varphi_2$).
The GRS and DPS $\eta_2$ signals showed the largest changes. 
The first plotted DWS data points show larger angles in yaw than in pitch. This is because the TMs were rotated by larger angles during the first realignment \cite{LPFdata22}. Note that the GRS angles were closer to zero which aligns to the observation that the TTL coupling was reduced due to this realignment step.

Next, we discuss which optical effects would change the three different angular readouts besides actual TM rotations.

Since the DWS angle is a measure of the relative angle between the measurement beam and the reference beam, it would also track if one of the beam directions were changed due to distortions in the optical setup, e.g.\ the fiber injector optical sub-assemblies (FIOSs) or the OB plate (further discussed in Sec.~\ref{sec:OBstability}).
In the analyzed timespans, the DWS readout was always used for the actuation control loops and kept stable. Hence, any differential beam tilt induced by distortions would not be visible in the plotted mean DWS readouts but yield a counter rotation of the TMs to preserve the DWS angles.
However, these TM rotations would be visible in the GRS and DPS readouts.

The GRS angles changed due to rotations of the TM relative to their housings. Therefore, the GRS is initially unaffected by beam tilts. However, due to mechanical stress or stress relieve, also the housings could slightly rotate, biasing the measurement.
Thus, the changes of the GRS readouts in Fig.~\ref{fig:TTLangles} could originate either from TM rotations commanded by the DFACS to counteract a changing DWS angle or from housing rotations. 

The measured DPS angles changed if either of the beams walked over the detector.
If not due to TM rotations, this beam walk also would have been induced by distortions of the optical setup. 
Tilts of the measurement beam would accumulate a translational offset of the beam's incident point at the detector. For small tilt angles, this offset would scale linearly with the magnitude of the tilt and the distance between the source of the rotation (e.g.\ the FIOS) and the detector or a counter-rotating component (e.g.\ a TM).
We will investigate the beam spot positions separately in Sec.~\ref{sec:CoeffStability_spot}.

In general, we cannot know which scenario yielded the angle changes just by looking at the readouts in Fig.~\ref{fig:TTLangles}.
For example, a simultaneous change of the GRS and the DPS angle, like in the case of $\eta_2$, could have been caused by a rotation of the TM2 housing in the opposite direction and an additional change of the height of the beams preserving the differential orientation of both interfering beams.
An equal up- or down-tilt of the beams could potentially lead to the latter.
On the other hand, we would see the same behavior if the differential angle between both interfering beams changed and the control mechanism applied a rotation of TM2 for compensation.
This shows that their comparison alone cannot provide a profound explanation of the underlying dynamics.
We will refine our analysis by investigating the beam spot positions next.

\subsection{Changes of the Normalized Spot Positions at the Detectors}
\label{sec:CoeffStability_spot}
Here, we discuss the spot positions of the various beams. These spot positions do not only help us to understand the DPS angle better but also yield an additional measure for the interpretation of the stability of the optical setup. 
In general, the shift of a beam's point of incidence on the photodiode surface (beam walk) changes the power distribution and, thus, the DPS angle.
The magnitude of this beam walk is linearly dependent on the distance between the detector and the origin of the beam tilt. 
In the following, we use this dependency to analyze the origin of the beam tilt. We know that a beam tilt at the FIOSs would have been visible in all four interferometers. Any (additional) beam tilt along the beam paths and after the first beamsplitter would only have been visible in some of the interferometers. E.g., tilts of the TMs would only yield a beam walk of the measurement beam in the x1- and x12-interferometers.
The computation of the DPS signal did not account for beam tilts other than from TM rotations. Thus, it would falsely indicated a test mass tilt, when in fact deformations caused a beam walk.

\begin{figure}
\flushright
  \includegraphics[scale=0.3]{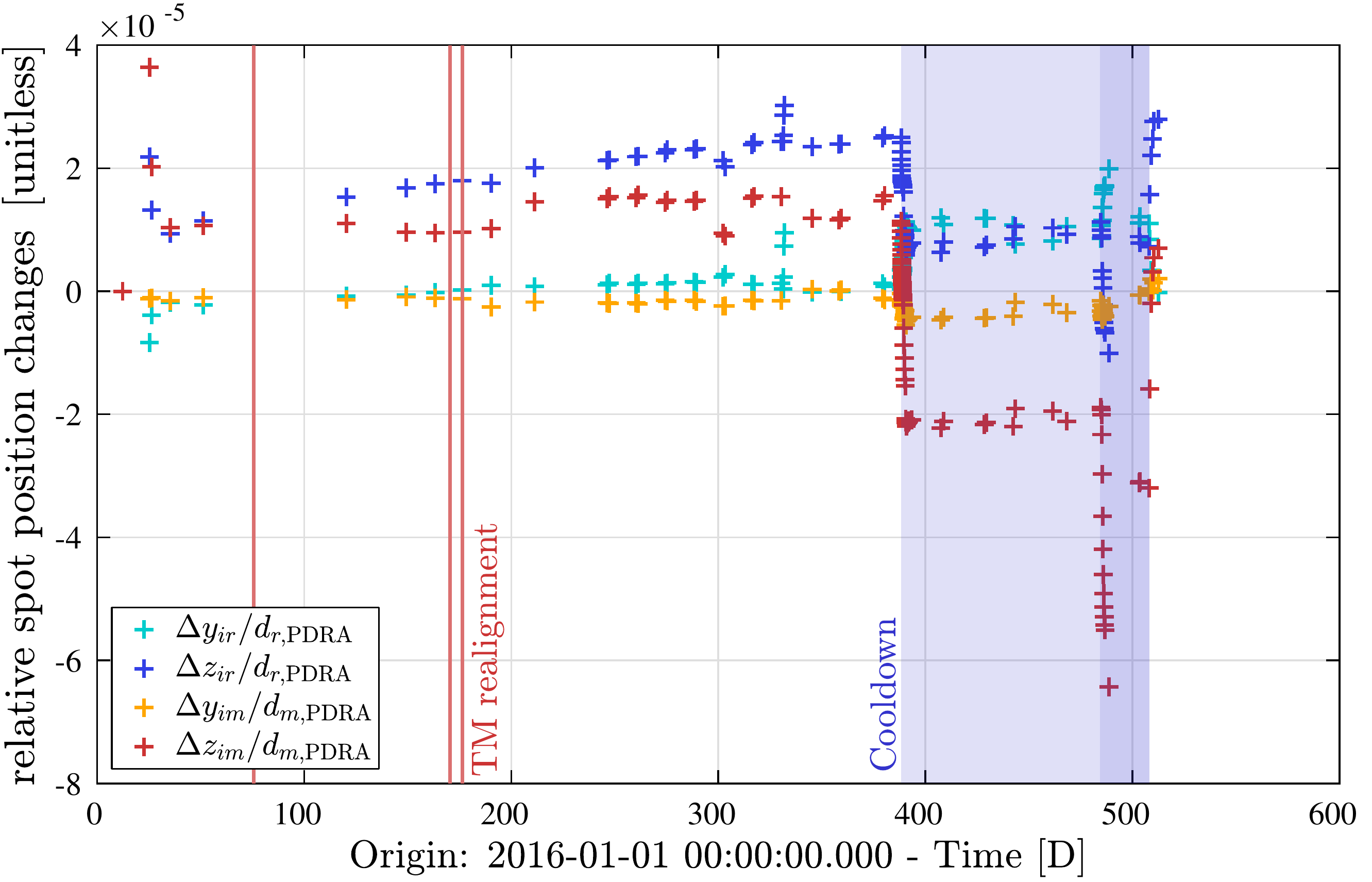} \\[0.5ex]
  \includegraphics[scale=0.3]{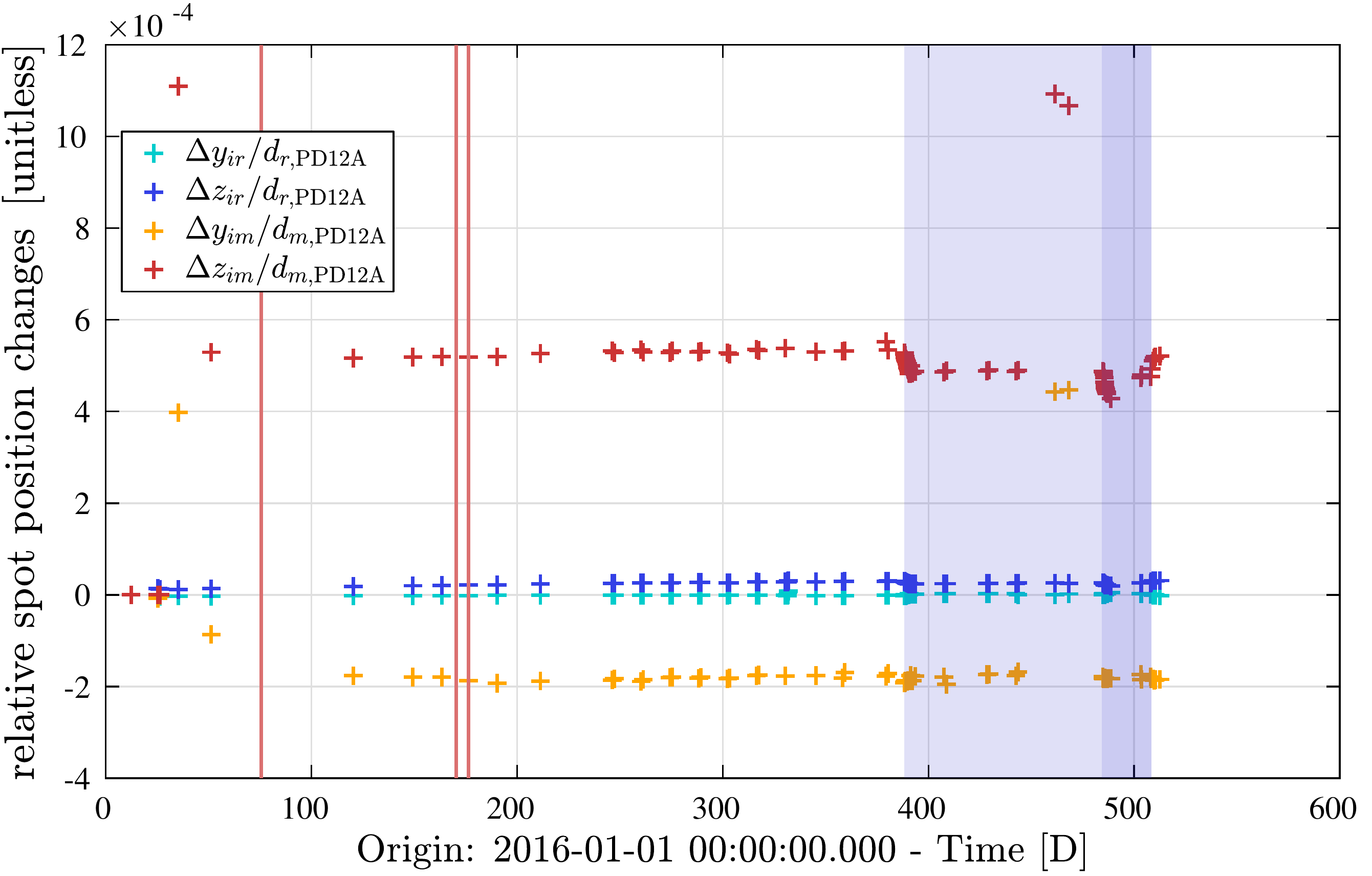}
  \caption{Relative spot position changes at the A-diodes of the xR- (top) and the x12-interferometer (bottom). 
  The horizontal beam walk is denoted with a $y$, and the vertical beam walk with a $z$. The indices $m$ and $r$ indicate the measurement (beam that is reflected at the TMs) and the reference beam. The spot position changes are divided by the pathlengths of the beams from the FIOSs to the respective detector.
  All beams' spot positions are affected by the cooldown. 
  Before the first TM realignment (i.e.\ during the commissioning phase), the measurement beam's spot position largely changed in the x12-interferometer.
  The large beam walk of the measurement beams' before the second phase of the cooldown is related to the change of the control mechanisms utilizing not the DWS but the GRS angular readouts at these times.}
\label{fig:TTLspotpositions}
\end{figure}

In Fig.~\ref{fig:TTLspotpositions}, we compare the spot positions for one out of the pair of the photodiodes (A-diodes) of the reference (xR) and the x12-interferometer.
These two interferometers were chosen here since the beams interfering in the xR-interferometer are independent of TM rotations, while one of the beams was reflected at both TMs in the case of the x12-interferometer.
Thus, a comparison of these two signals allows a differentiation between TM tilts and beam tilt at the FIOS output.
The hot redundant B-diodes yield a very similar result, which is for completeness shown in App.~\ref{sec:CoeffStability_spotApp}.

Note that the plotted spot positions stem from dedicated measurements \cite{Armano2022_OMS} only possible if one of the beams were switched off. Therefore, the measurements times corresponding to the data points in Fig.~\ref{fig:TTLspotpositions} are not identical to the URLA timespans listed in Tab.~\ref{tab:timespans}. However, we can compare the trends of the spot positions with the angular readouts or coupling coefficients.

For better comparison, we did not plot the absolute measurement of the spot positions in Fig.~\ref{fig:TTLspotpositions} but their difference from the first measurement. 
Additionally, we divided the measurements by the optical pathlength of the respective beam in between its origin (FIOS) and the detector.
This allows us to compare the readouts from the two interferometers independently of the propagation distance of the beams.
So, the beam walk originating from the FIOS would be of the same magnitude in both plots, while the beam walk of any other origin would be different for both interferometers.

In general, we see that the spot positions in both interferometers are affected by the cooldown. This effect is stronger in the vertical direction than in the horizontal.
Comparing the changes of the scaled vertical beam walk of the measurement beam (red crosses), we find that it is slightly larger in the x12-interferometer (about 50$\times10^{-6}$) than in the xR-interferometer (less than 40$\times10^{-6}$). 
From the relationships that we explained in the previous paragraphs, we interpret the common beam walk as a distortion of the optical components in the early beam path, e.g.\ a beam tilt at their origin (Sec.~\ref{sec:OBstability_FIOS}).
The large magnitude of the changes makes a beam tilt very likely as the beam offset would increase along the beam path.
Furthermore, the measurement beam must have undergone an additional change, e.g.\ due to TM rotations, yielding the difference between both interferometer measurements.

In comparison, the vertical beam walk of the reference beam (blue crosses) is affected less by the temperature changes. 
This can be seen in the xR- and the x12-interferometer, indicating a partially common reason for the spot position changes.
Therefore, we expect a tilt of the reference beam in its early beam path. Presumably, this tilt was smaller than for the measurement beam.

The horizontal beam walk of both the measurement and the reference beam is small in both interferometers. We conclude that the cooldown affected the optical setup marginally in the horizontal plane.

\subsection{Interpretation of the Long-Term Data Analyses}
\label{sec:CoeffStability_interpretation}

So far, we have described the long-term behavior of the fitted TTL coupling coefficients, the angular readouts for the same timespans, and the change of the normalized spot positions separately.
The angle and spot position measurements yield no definite interpretation of the source of their changes. 
We present here that the TTL coupling analysis can provide an additional measure for the setup stability.

Fig.~\ref{fig:TTLcoeffs} shows that the $C_\eta$ and the $C_z$ coefficients changed most significantly due to the cooldown.
It was discussed in \cite{LPFana22} that the pitch coefficient $C_\eta$ is sensitive to changes in the beams' angular alignment at their source and the point of reflection at the TMs. Both effects couple very little with the TM tilts and would mostly couple into the offset $C_{\eta,0}$ in Eq.~\eqref{eq:anaCeta}.
However, the $C_z$ coefficient almost only depends on the relative angular alignment of the TMs \cite{LPFana22}.
Hence its change in response to the cooldown must originate from real TM rotations. 
Using Eq.~\eqref{eq:anaCeta}, we can interpret the jumps of the coefficient due to a temperature change of \SI{10}{K} (Fig.~\ref{fig:TTLcoeffs}) by a differential rotation angle of the TMs by 30\,$\upmu$rad. 

Principally, we have to discuss the changes in the $C_y$ coefficient analogously.
However, its changes are comparatively small such that the observed change might be dominated by uncertainties of the fit \cite{LPFdata22} or correlations to the $C_{y,s}$ coefficient, see App.~\ref{sec:CoeffStability_CoeffsApp}. 
However, we can still conclude that the beams were significantly more stable in yaw than in pitch.
This conclusion is consistent with the overall smaller changes of the GRS and DPS yaw angles (Fig.~\ref{fig:TTLangles} and the normalized horizontal spot position changes (Fig.~\ref{fig:TTLspotpositions}). 

Since the DWS angles did not change due to the cooldown, the pitch rotations of the TMs must have been accompanied by an opposite beam rotation from another source, see Sec.~\ref{sec:OBstability}.
If any strain or warp of the setup would have led to a differential angle between the two interfering beams, the TMs would have been rotated to preserve the predefined DWS angles.
Both sources of beam rotations would have changed the DWS and the DPS readout. 
However, the GRS angles would only show the applied rotations of the TMs with respect to their housings.
For this reason, we use the GRS angles to further analyse the coefficient changes, despite the fact that we cannot exclude an additional deformation of the EHs.

We insert the changes of the GRS angles due to the first (23.\,Jan.\ 2017) and second (29.\,Apr.\ 2017) temperature decrease during the cooldown experiment and the final heat-up into the analytical equations for the $C_z$ (Eq.~\eqref{eq:anaCz}) and the $C_y$ (Eq.~\eqref{eq:anaCy}) coefficient. 
These analytically computed coefficient changes are added to the fit result for the previous timespan. 
The resulting analytical coefficients are compared to the fitted ones in Tab.~\ref{tab:CD_coeffs}.

\begin{table}
\begin{tabular}{lcrrcrr}
\toprule
coefficient [$10^{-6}$] &\quad\ & $C_y^\text{fit}$ & $C_y^\text{ana}$
                        &\quad\ & $C_z^\text{fit}$ & $C_z^\text{ana}$ \\          
\midrule
before cooldown        &\ & 1.5 &   --  &\ & 15.8 &   -- \\
1st phase of cooldown  &\ & 2.9 &  4.2  &\ & 44.8 & 46.3 \\
2nd phase of cooldown  &\ & 7.9 & -0.3  &\ & 79.3 & 84.5 \\
after cooldown         &\ & 2.1 & 10.3  &\ & 17.4 & 15.1 \\
\bottomrule
\end{tabular}
\caption{Comparison of the fitted and analytically predicted lateral coupling coefficients before, during and after the cooldown. 
In the case of the fitted cooldown-coefficients, we show the mean coefficient of the computations within these cooldown phases.
The analytical coefficients are computed via Eq.~\eqref{eq:anaCy} and Eq.~\eqref{eq:anaCz} using the previous fit result as constant offset and inserting the mean angular GRS readouts.}
\label{tab:CD_coeffs}
\end{table}

For the $C_z$ coefficient, the GRS angles can explain almost the full changes. 
The residual differences of both sets of coefficients can partially originate from uncertainties of the fit. It has been shown in \cite{LPFdata22} that the fitted $C_z$ coefficients within a noise run during the cooldown have an error (root-mean-square) of 0.55$\times10^{-6}$.
Further deviations might originate from an additional rotation of one or both EHs, which couple into the GRS readouts.

In the case of the $C_y$ coefficients, the analytical prediction using the GRS-angles does not match well.
Even though the absolute deviations are only about 50\,\% larger than in the case of the $C_z$, the mismatch is significantly more apparent due to the smaller coefficient values.
Also, uncertainties of the fit result and, during the second phase of the cooldown, correlations with the stiffness coefficient $C_{y,s}$ (see App.~\ref{sec:CoeffStability_CoeffsApp}) could lead to the deviation of the fitted $C_y$.
Furthermore, we expect that the EHs rotated in yaw by a few micro-radians due to the temperature changes and thereby changed the GRS yaw angles.

\section{Discussion of the Optical Setup Stability}
\label{sec:OBstability}

We have argued in the previous section that the TMs rotated in pitch due to the cooldown of the S/C. 
At the same time, the DWS angular readouts did not change (Fig.~\ref{fig:TTLangles}).
Since a differential rotation of the TMs would normally also affect the DWS measurements, the differential beam alignment must have been changed by another mechanism having the inverse effect.
We hypothesize that this inverse effect was a temperature-dependent rotation of the rays on the optical bench relating to stresses or relaxations in the setup.
This rotation would have been counteracted by a tilt of the TMs which then yielded the changes in the GRS readout and the TTL coupling coefficients. 

In the following, we will investigate two types of optical distortions, which both could have originated from temperature changes and were already investigated prior to the LPF mission: 
A beam tilt at the FIOSs and a bending of the OB baseplate \cite{Killow2016,Armano2022_OMS}.

\subsection{Beam tilt at the FIOSs}
\label{sec:OBstability_FIOS}

A potential temperature-related beam tilt at the FIOSs has been studied in preparation for the LPF mission.
It was found that the beam could tilt by up to 3\,$\upmu$rad/K in pitch \cite{Killow2016}.
While this distortion was negligible most of the mission time, it might have become relevant for the large temperature decrease at the end of January 2017.
For the temperature decrease of \SI{10}{K}, we would find a beam down-tilt of up to 30\,$\upmu$rad.

A beam tilt at its source, the FIOS, would have changed its spot positions in all interferometers. 
This temperature-induced effect would be expected to affect the FIOSs of both beams similarly since they are of equivalent design and have been placed close to each other on the OB.
In Fig.~\ref{fig:TTLspotpositions}, we plotted beam spot positions normalized by the propagation distances from the FIOS to the detector. If the spot position changes were only due to tilting of the FIOS, in a small-angle approximation, these readouts would correspond to the tilt angle.

Let us now compare the out-of-plane (pitch) tilts in Fig.~\ref{fig:TTLspotpositions} with the spot position changes of the measurement beam in the reference interferometer (upper plot in Fig.~\ref{fig:TTLspotpositions}).
The magnitude of the angular tilt (30\,$\upmu$rad for temperature changes of \SI{10}{K}) is slightly smaller than the normalized spot position changes (approximately 40\,$\upmu$rad assuming that the FIOS would be the only source of this tilt).
Moreover, we find a down-tilt of the reference beam of roughly half the magnitude as in the case of the measurement beam.

We cross-checked our derivations with simulations using the optical simulator IfoCAD \cite{Wanner2012,IfoCAD,Kochkina2013} and the LPF in-flight model \cite{TNoptocad}. 
In these simulations, we find a normalized spot position change of almost 3$\times10^{-5}$ in negative direction for the measurement beam assuming a down-tilt of 30\,$\upmu$rad (in PDRA). For the reference beam, the position change is of approximately half this magnitude for a down-tilt of 15\,$\upmu$rad. 
These computations only slightly underestimate the measured normalized spot position changes shown in Fig.~\ref{fig:TTLspotpositions} (about 4$\times10^{-5}$ and 2$\times10^{-5}$ in the upper plot).

However, these beam tilts would only partially be measurable by the DWS of the x12-interferometer tracking the change of the differential angular beam alignment. 
To counteract the differential beam alignment of approximately 15\,$\upmu$rad, a TM1 rotation would have been applied by half this angle, i.e. $\eta_1\approx 7.5\,\upmu$rad. 
(For geometrical reasons, the reflected beam rotates by twice the TM angle.)
The TM2 alignment would remain unchanged.
This TM realignment is significantly smaller than the differential angle of 30\,$\upmu$rad needed to explain the change of the $C_z$ coefficient (compare Sec.~\ref{sec:CoeffStability_interpretation}). 

In conclusion, a beam tilt at the FIOSs due to the cooldown was likely but cannot explain the change of the TTL coupling coefficient $C_z$ alone. 
A second type of distortion must have led to the large coefficient variation.

\subsection{Bending of the Optical Bench}
\label{sec:OBstability_OB}

\begin{figure}
\centering
\includegraphics[width=0.8\columnwidth]{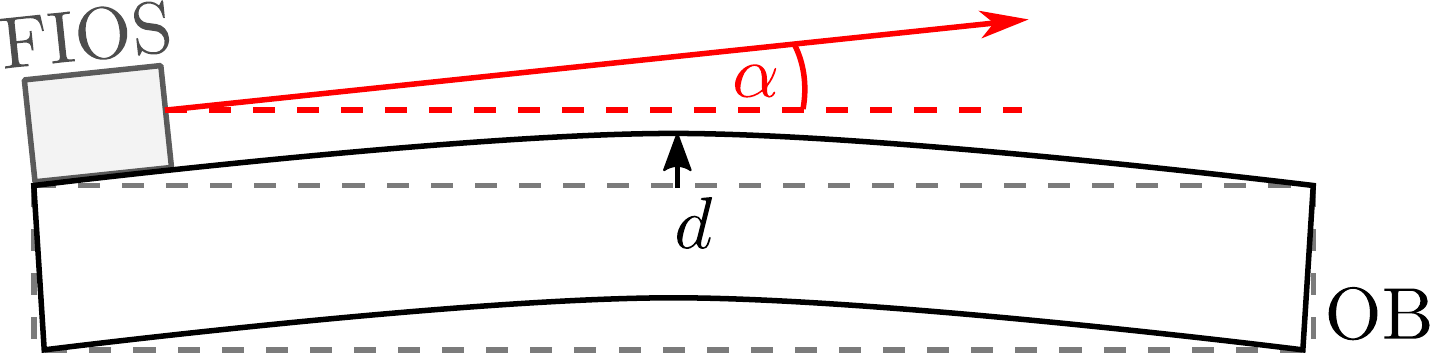}
\caption{Beam tilt due to a bending optical bench.
In this figure, the angle $\alpha$ is the pitch angle of the beam at its origin in a global coordinate system for an OB bulge of height $d$.}
\label{fig:OBbending}
\end{figure}

\begin{table}
\caption{Changes of the pitch angles of both TMs ($\eta_1$, $\eta_2$, $\Delta\eta=\eta_1-\eta_2$) due to the cooldown. 
  The first two columns show the differential tilt required to explain the TTL coupling coefficient $C_z$ change and the approximate GRS angle readout taken from Fig.~\ref{fig:TTLangles}. The first number corresponds to the changes at the beginning of the cooldown, the number in the brackets are the additional changes due to the second cooldown phase with colder temperatures. Note that we can only derive the differential angle change from the change of $C_z$, but not the actual angles.
  The IfoCAD simulation results show the TM realignments necessary to compensate for the computed differential angles at the x1- and x12-interferometer. We consider three different scenarios: 
  (a) Beam tilt at the FIOSs ($\eta_m=-30\upmu$rad, $\eta_r=-15\upmu$rad), see Sec.~\ref{sec:OBstability_FIOS};
  (b) OB bending about the $y$-axis;
  (c) OB bending about the $x$-axis.
  Gray entries: Angles computed via the other numbers in the same column. 
  }  
\begin{tabular}{lc|ccc|cccc}
\toprule
 & & \multicolumn{2}{c}{LPF data analysis} 
 & & \multicolumn{3}{c}{IfoCAD simulation} \\
                         & & $\Delta C_z$     & GRS        
                         & & FIOS      & $y$-axis     & $x$-axis \\
\midrule
$\eta_1$ [$\upmu$rad]    & & --        &   0 (+15)  
                         & & +7.5      & +14          & $-$18 \\
$\eta_2$ [$\upmu$rad]    & & --        & $-$30 ($-$25)  
                         & & \grey{0}  & \grey{$-$12} & \grey{$-$32} \\
$\Delta\eta$ [$\upmu$rad]& & +30 (+30) & \grey{+30 (+40)}
                         & & +7.5      & +36          & +16 \\
\bottomrule
\end{tabular}
\label{tab:OBdistortion}
\end{table}

A second distortion measured prior to the LPF launch was a distortion of the OB baseplate, as shown in Fig.~\ref{fig:OBbending}.
Due to the stresses (or the relaxations) of its mounting, the baseplate could bend. 
A distortion of up to 1.7\,$\upmu$m ($d$ in Fig.~\ref{fig:OBbending}) has been measured in the laboratory \cite{Armano2022_OMS} and is plausible for extreme temperature changes during the cooldown.  

For such a deformation, we compute a beam tilt at the FIOS between 25 and 34\,$\upmu$rad, depending on how the FIOS responded to the baseplate curving. 
We assumed a perfect cylindrical curving of the OB in our computation.
Since the mirrors and beamsplitters were also tilted due to the distorted OB, the beam's pitch angle changed with each reflection.  
This was the case for both the measurement and the reference beam.
However, their alignment before interference depended on the individual number and position of the reflecting components along their path. 

We simulated the expected beam tilts at the diodes of the x12-interferometer with IfoCAD for three different scenarios.
First, we assumed only a beam tilt at the FIOS as discussed in Sec.~\ref{sec:OBstability_FIOS}. Second and third, we chose a cylindrical bending of the optical bench along the $y$- and then the $x$-axis (cf.\ coordinate system in Fig.~\ref{fig:LPFscetch}).
A bending along the $y$-axis could be explained by thermal stress on the OB mounting, which was supporting the sides of the OB not occupied by the EHs (compare Fig.~\ref{fig:LPFscetch}). 
For comparison, we also investigate the bending along the orthogonal $x$-axis.
We computed the differential angle of the measurement and reference beam at the x1- and x12-interferometers. From these numbers, we derived the TM realignment angles required to counteract these differential angles. The results are summarized in Tab.~\ref{tab:OBdistortion}.

It becomes evident that none of these scenarios perfectly matches the differential TM tilt required to explain the TTL coupling coefficient change or the measured GRS angles (also shown in Tab.~\ref{tab:OBdistortion}).
However, we see that OB deformations would have caused TM realignments in the same order of magnitude as the angles required to explain the TTL coupling coefficient changes. Therefore, they could explain the significant tilts of the TMs during the cooldown. 
If considering a simultaneous deformation of the OB and beam tilt at the FIOSs, an OB bending about the $y$-axis yields angles closer to the GRS readout than in the case of a bending about the $x$-axis. 
As described above, we consider this scenario to be more feasible.

Mind that these computations all assume a perfectly cylindrical bending of the bench, i.e.\ no unsymmetrical curving, which likely had not occurred like this.
Also, the FIOSs' tilt was computed from the beams' spot positions in Sec.~\ref{sec:OBstability_FIOS} under the assumption that the spot position changes only occurred from the FIOSs' tilt. 
This approach cannot be used for the computation of the FIOSs' angles if we consider the OB bending as well since that also changes the spot positions. 

In summary, a beam tilt at the FIOSs together with an unsymmetrical distortion of the OB baseplate could explain the observed TTL coupling changes.
The bending of the OB, despite being very small, has a considerable effect on the differential beam angles at the detectors and, therefore, could have caused significant TM realignments.

\section{Summary}
\label{sec:summary}

We have shown in this paper that the TTL coupling coefficients were not entirely stable during the LPF mission. 
Besides the planned coupling minimization due to TM realignment, the coupling coefficients drifted (by less than 1\,$\upmu$m/rad or 6$\times10^{-6}$ within 100\,days) and showed a strong temperature dependency. 
Particularly the out-of-plane coefficients $C_\eta$ and $C_z$ changed significantly in response to temperature variations (8\,$\upmu$m/rad/K and 30$\times10^{-6}$/K).

Based on the analytical TTL coupling model presented in \cite{LPFana22}, we concluded that the observed coefficient changes during the S/C cooldown, which started in late 2016, were caused by pitch rotations of the TMs.
No such rotation was visible in the long-term DWS readouts, which remained constant.
However, we have shown in this paper that (temperature-driven) beam tilts at the FIOS, and the bending of the OB baseplate could have caused significant differential beam angles at the detectors. 
These beam misalignments would have been corrected during the mission by the feed-back loops. The counteracting rotations of the TMs were partially visible in the GRS angular readouts. 
We used the information of the coupling coefficient changes to deduce information about the stability of the OB. By this means,
we have shown, that the magnitude of the observed coupling coefficients could be explained by OB deformations.
An additional contribution due to a beam tilt at their origin (FIOS) is likely.

The presented investigation of the stability of the optical system was only possible with an analytical TTL coupling model.
Therefore, this model presents a useful additional tool for this type of investigation.
It is unclear whether a comparable analysis can be repeated for LISA even if we had an analytical model.
The here presented modeling mostly relied on the equations for lateral jitter coupling. This coupling is expected to be much smaller in LISA \cite{Wanner2024}. 
For angular coupling coefficients, such investigations are more complicated since they depend on a variety of mechanisms \cite{LPFana22}.
In the analytical model derivations two Gaussian beams were assumed \cite{LPFana22}, while there will be inevitable beam clipping in LISA. Other characteristic differences between the two missions are the LISA telescopes and imaging optics.

Also the observed long-term coefficient drifts will be different in LISA. 
While the angular coefficient drifts stayed below 1\,$\upmu$mm/rad in 100 days, there is an expected change of 0.15\,mm/rad in one day \cite{Paczkowski2022}.

Mind that the strong deviations of the coefficients (and hence the distortion of the system) mostly originated from an intentionally big decrease of the temperature within the S/C.
Hence, the described deformation of the optical system is not expected in LISA. Furthermore, the FIOS design has been updated (all fused silica constructions, which have a lower coefficient of thermal expansion), reducing beam tilts from this origin.

\begin{acknowledgments}
    \label{S:acc}
    This work has been made possible by the LISA Pathfinder mission, which is part of the space-science programme of the European Space Agency.
    The Albert Einstein Institute acknowledges the support of the German Space Agency, DLR. 
    The work is supported by the Federal Ministry for Economic Affairs and Climate Action based on a resolution of the German Bundestag (FKZ 50OQ0501, FKZ 50OQ1601 and FKZ 50OQ1801). We also acknowledge the support by the Deutsche Forschungsgemeinschaft (DFG, German Research Foundation) under Germany’s Excellence Strategy (EXC-2123 QuantumFrontiers, project ID 390837967).
    The French contribution has been supported by the CNES (Accord Specific de projet CNES 1316634/CNRS 103747), the CNRS, the Observatoire de Paris and the University Paris-Diderot. E. Plagnol and H. Inchauspé would also like to acknowledge the financial support of the UnivEarthS Labex program at Sorbonne Paris Cité (ANR-10-LABX-0023 and ANR-11-IDEX-0005-02).
    The Italian contribution has been supported by ASI (grant n.2017-29-H.1-2020 ``Attività per la fase A della missione LISA'') and Istituto Nazionale di Fisica Nucleare.
    The Spanish contribution has been supported by contracts AYA2010-15709 (MICINN), ESP2013-47637-P, ESP2015-67234-P, and ESP2017-90084-P (MINECO). Support from AGAUR (Generalitat de Catalunya) contract 2017-SGR-1469 is also acknowledged.
    M. Nofrarias acknowledges support from Fundacion General CSIC (Programa ComFuturo). 
    F. Rivas acknowledges an FPI contract from MINECO.
    The Swiss contribution acknowledges the support of the ETH Research Grant ETH-05 16-2 and the Swiss Space Office (SSO) via the PRODEX Programme of ESA. L. Ferraioli is supported by the Swiss National Science Foundation.
    The UK groups wish to acknowledge support from the United Kingdom Space Agency (UKSA), the Scottish Universities Physics Alliance (SUPA), the University of Glasgow, the University of Birmingham and Imperial College London.
    J. I. Thorpe and J. Slutsky acknowledge the support of the US National Aeronautics and Space Administration (NASA).
    N. Korsakova would like to thank for the support from the CNES Fellowship.
    The LISA Pathfinder collaboration would like to acknowledge Prof. Pierre Binetruy (deceased 30 March 2017), Prof. José Alberto Lobo (deceased 30 September 2012) and Lluis Gesa Bote (deceased 29 May 2020) for their remarkable contribution to the LISA Pathfinder science. 
\end{acknowledgments}

\appendix

\section{Timespans for Long-Term Analysis}
\label{sec:timespans}

The TTL coefficients are sensitive to changes in the setup and the environment. 
Therefore, we only investigate here the coefficients evaluated for noise runs with very low actuation (URLA segments).
Among these, we considered only time segments longer than 40000\,s to keep the numerical variations of the coefficients low.
Three additional time segments have been discarded due to large bias voltages or TM charge concerns.
The alignment stability in the remaining timespans has been controlled utilizing the DWS readout, except for two cases during the second phase of the cooldown, where the GRS readout has been used. 
This change in the control mechanism yielded a realignment of the TMs and hence the coupling coefficients. 
Therefore, we also discarded these two segments from our analysis.
All other timespans are summarized in Tab.~\ref{tab:timespans}.

\begin{table}
  \centering
  \begin{tabular}{rrll}
    \toprule 
    \# & run\hspace{1ex}   & start time         & end time             \\
    \  & index\hspace{1ex} & [dd:mm:yyyy hh:mm] & [dd:mm:yyyy hh:mm]   \\
    \toprule
    1  &  6\hspace{2ex} & 21.03.2016 02:00  & 26.03.2016 08:00  \\
    2  &  9\hspace{2ex} & 04.04.2016 00:00  & 14.04.2016 08:00  \\
    3  & 12\hspace{2ex} & 26.04.2016 08:00  & 28.04.2016 08:00  \\
    4  & 13\hspace{2ex} & 01.05.2016 08:05  & 02.05.2016 23:55  \\
    5  & 16\hspace{2ex} & 13.05.2016 08:30  & 14.05.2016 08:00  \\
    6  & 17\hspace{2ex} & 16.05.2016 00:00  & 19.05.2016 05:00  \\
    7  & 31\hspace{2ex} & 06.06.2016 11:05  & 09.06.2016 08:00  \\
    8  & 39\hspace{2ex} & 19.06.2016 13:00  & 24.06.2016 08:00  \\
    9  & 40\hspace{2ex} & 10.07.2016 08:40  & 11.07.2016 09:55  \\
    10 & 41\hspace{2ex} & 11.07.2016 11:40  & 12.07.2016 09:55  \\
    11 & 42\hspace{2ex} & 17.07.2016 22:00  & 20.07.2016 06:00  \\
    12 & 43\hspace{2ex} & 24.07.2016 13:00  & 30.07.2016 00:00  \\
    13 & 44\hspace{2ex} & 31.07.2016 11:40  & 02.08.2016 06:00  \\
    14 & 45\hspace{2ex} & 07.08.2016 10:20  & 08.08.2016 04:20  \\
    15 & 53\hspace{2ex} & 19.09.2016 02:32  & 21.09.2016 13:00  \\
    16 & 53\hspace{2ex} & 21.09.2016 13:45  & 22.09.2016 06:00  \\
    17 & 54\hspace{2ex} & 28.09.2016 13:35  & 01.10.2016 08:00  \\
    18 & 56\hspace{2ex} & 05.10.2016 17:25  & 07.10.2016 00:49  \\
    19 & 56\hspace{2ex} & 07.10.2016 02:15  & 08.10.2016 07:50  \\
    20 & 58\hspace{2ex} & 07.11.2016 21:30  & 12.11.2016 08:00  \\
    21 & 59\hspace{2ex} & 16.11.2016 11:05  & 26.11.2016 08:00  \\
    22 & 61\hspace{2ex} & 26.12.2016 08:00  & 13.01.2017 19:58  \\
    23 & 63\hspace{2ex} & 27.01.2017 18:45  & 28.01.2017 08:00  \\
    24 & 64\hspace{2ex} & 02.02.2017 07:55  & 02.02.2017 20:20  \\
    25 & 66\hspace{2ex} & 13.02.2017 14:30  & 03.02.2017 21:50:19  \\
    26 & 67\hspace{2ex} & 09.03.2017 19:20  & 14.03.2017 09:40  \\
    27 & 68\hspace{2ex} & 14.03.2017 09:00  & 17.03.2017 00:30  \\
    28 & 71\hspace{2ex} & 03.05.2017 23:30  & 09.05.2017 14:00  \\
    29 & 72\hspace{2ex} & 10.05.2017 11:11:20  & 12.05.2017 12:02:07  \\
    30 & 74\hspace{2ex} & 18.05.2017 18:24:46  & 23.05.2017 02:00  \\
    31 & 75\hspace{2ex} & 28.05.2017 13:41  & 05.06.2017 15:04:40  \\
    32 & 76\hspace{2ex} & 08.06.2017 12:00:45  & 17.06.2017 02:56  \\
    33 & 80\hspace{2ex} & 15.07.2017 00:50  & 17.07.2017 13:45  \\
        \bottomrule
    \end{tabular}
    \caption{URLA timespans used in this work for the fit of the TTL coupling coefficients (long-term analysis) and the corresponding LPF run index. The actuation on TM2 was reduced in the URLA configurations compared to the nominal settings. Times are given in UTC.}
    \label{tab:timespans}
\end{table}

\section{Long-Term Stability of the Stiffness Terms and the $o_1$-Residual}
\label{sec:CoeffStability_CoeffsApp}

\begin{figure}
  \flushright
  \includegraphics[scale=0.295]{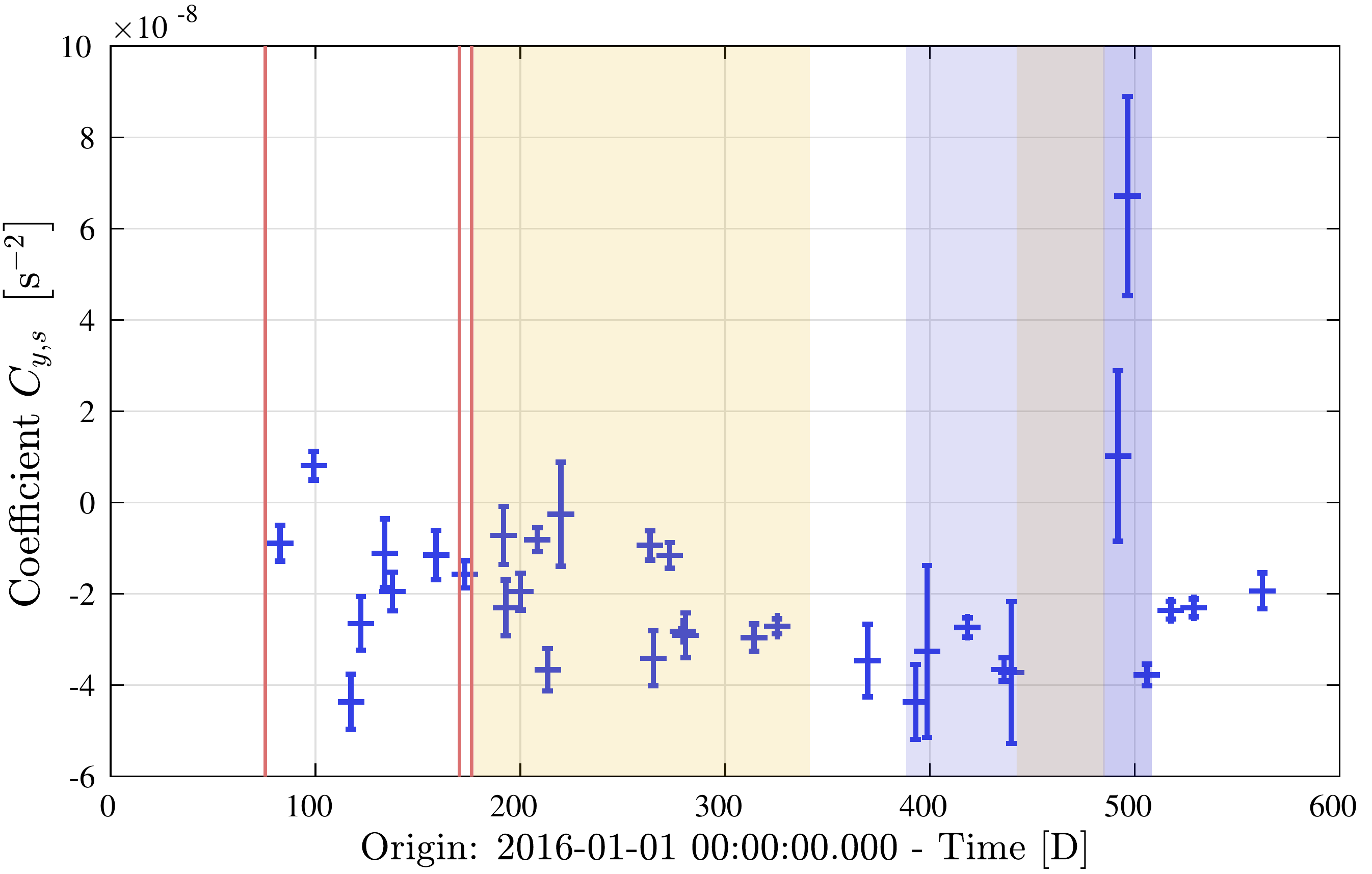} \\[1ex]
  \includegraphics[scale=0.295]{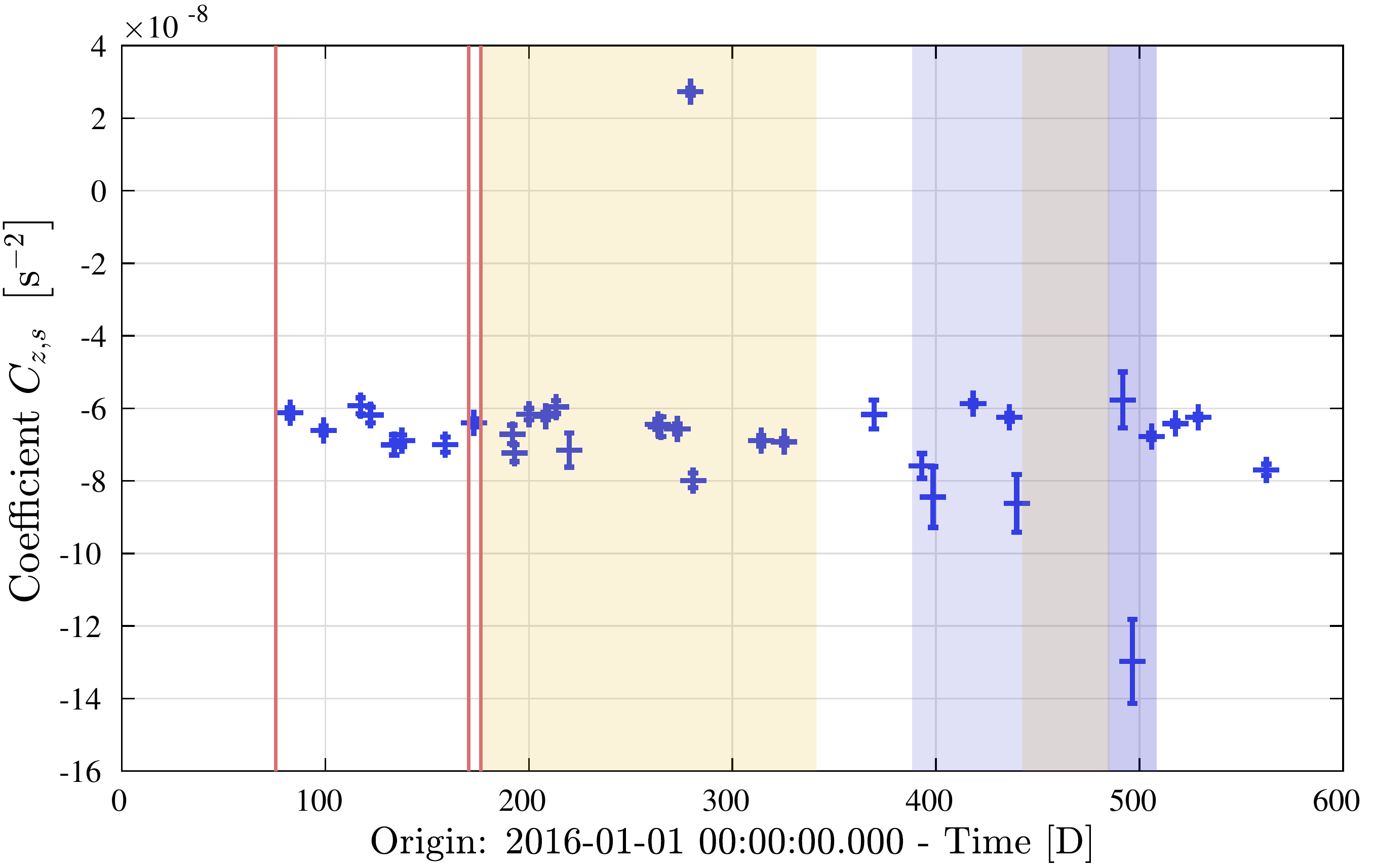} \\[1ex]
  \includegraphics[scale=0.295]{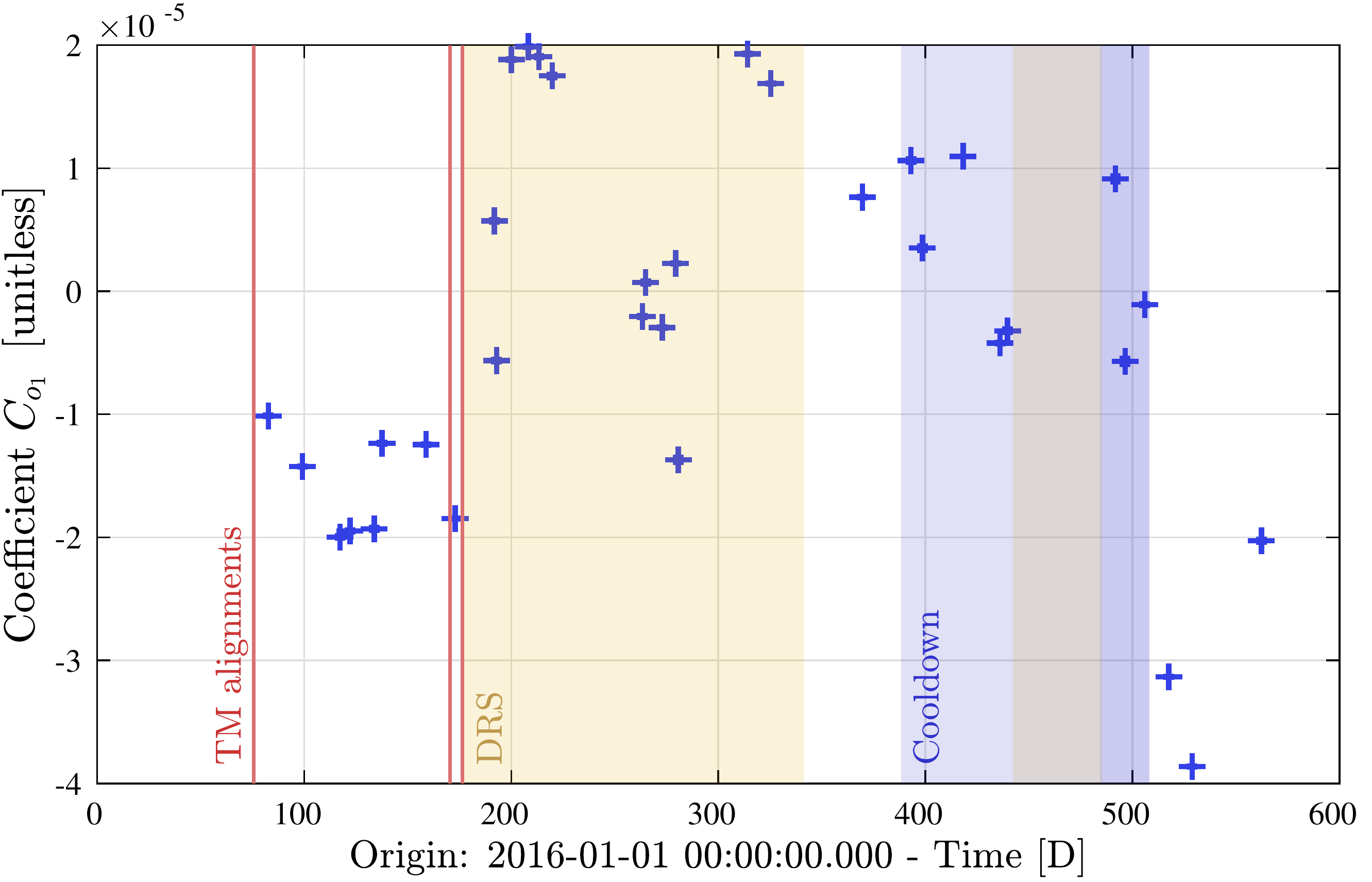}
  \caption{Long-term behaviour of the fitted TTL coefficients for the stiffness and the $o_1$-contributions. These coefficients complete the set shown in Fig.~\ref{fig:TTLcoeffs}.}
\label{fig:TTLcoeffs_stiff}
\end{figure}

In Fig.~\ref{fig:TTLcoeffs}, we have only shown the TTL coupling coefficients for the lateral and angular accelerations, which are the dominant TTL noise contributors.
For completeness, we show in Fig.~\ref{fig:TTLcoeffs_stiff} the other three coefficients that have simultaneously been fitted: the cross-coupling of lateral TM displacements via stiffnesses and the residual S/C jitter along the $x$-axis.

All three coefficients show no significant drift or response to the cooldown.
The stiffness terms are partially highly uncertain.
Particularly, the larger outliers during the second phase of the cooldown could indicate a change of the fit results for the lateral coupling coefficients due to correlations.
However, the relative changes of the larger lateral coupling coefficients would be small.

\section{Stability of the Spot Positions at the B-Diodes}
\label{sec:CoeffStability_spotApp}

\begin{figure}
  \flushright
  \includegraphics[scale=0.3]{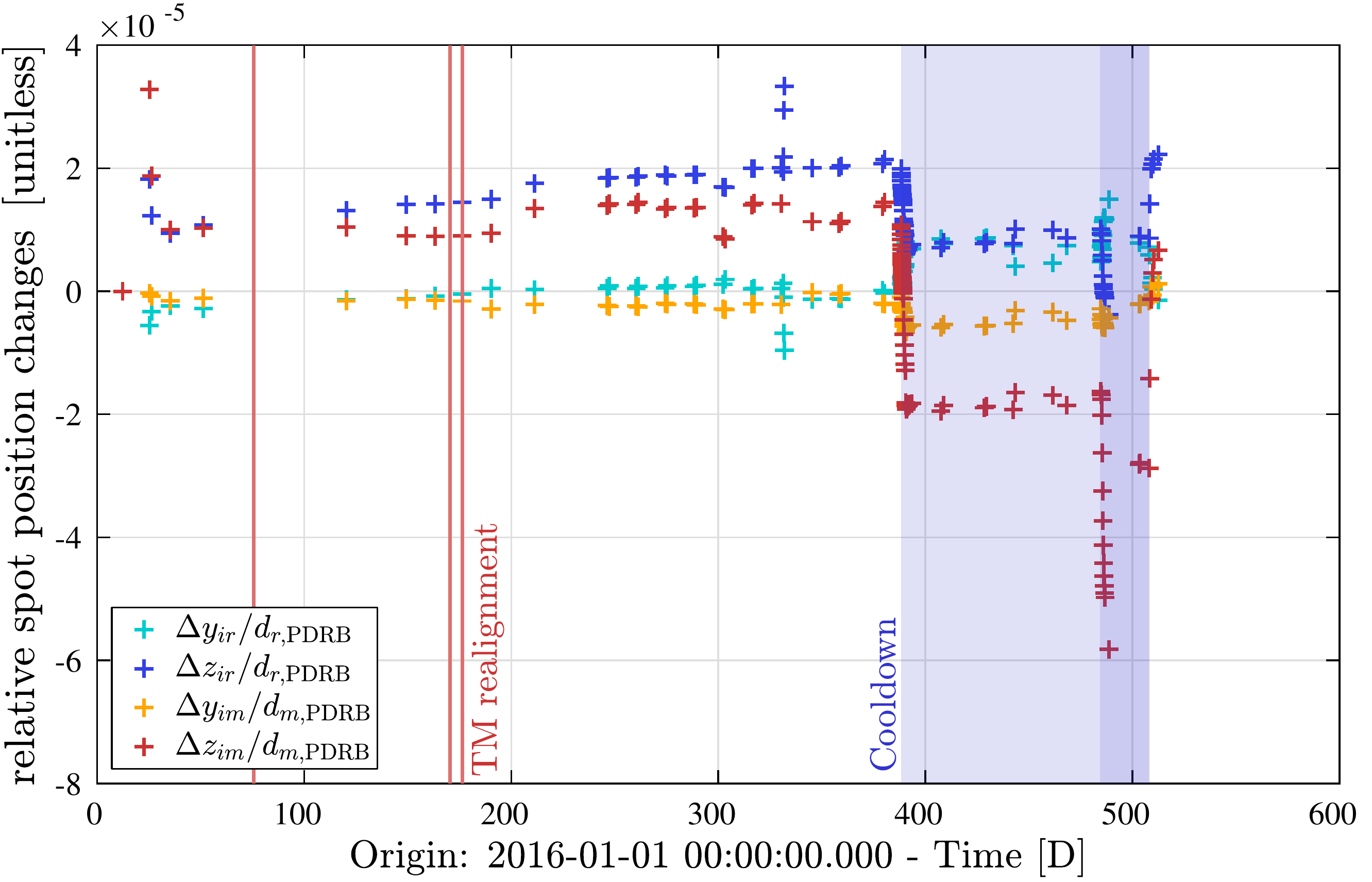} \\[0.5ex]
  \includegraphics[scale=0.3]{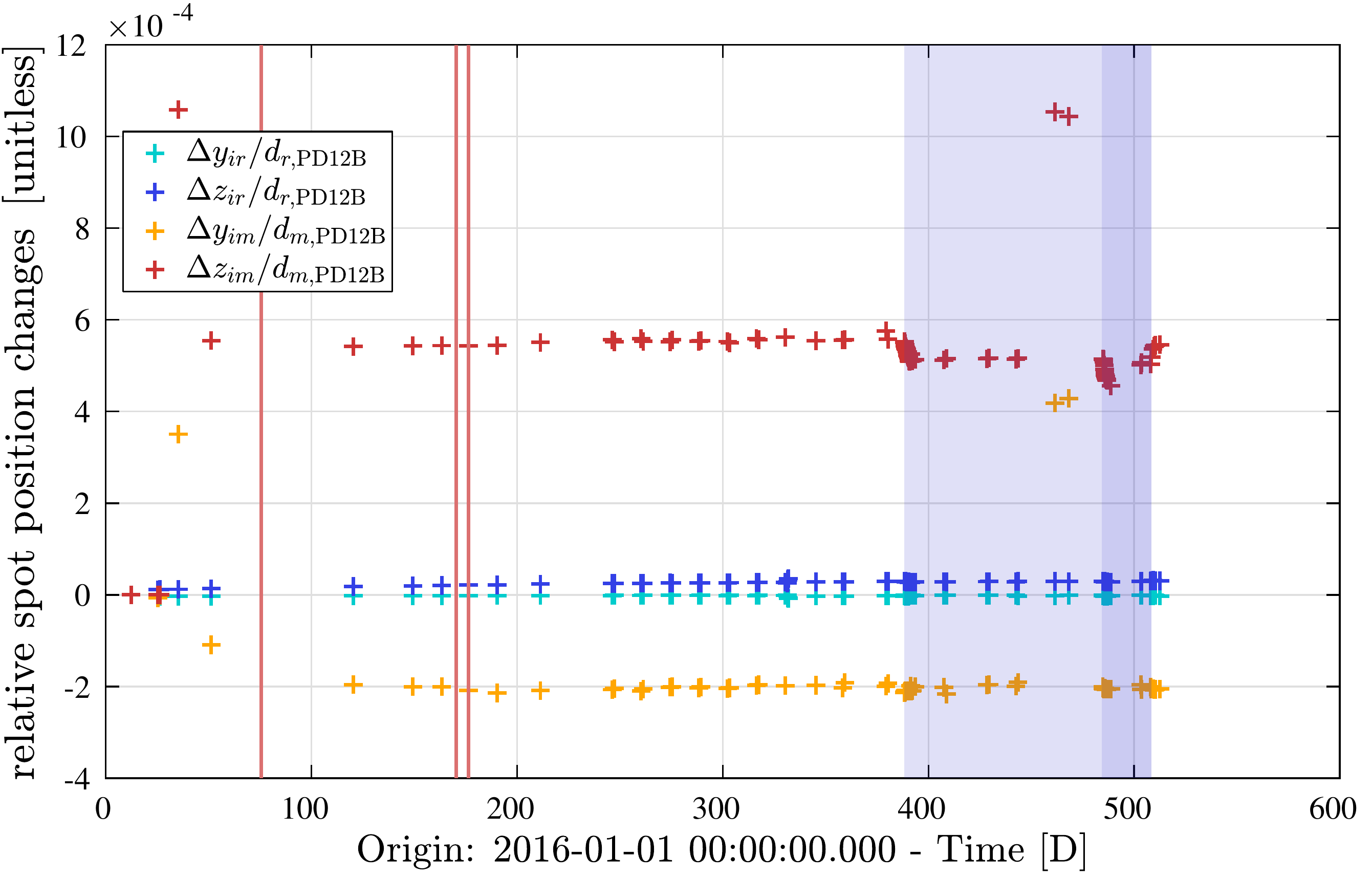}
  \caption{Relative spot position changes at the B-Diodes of the xR- (top) and the x12-interferometer (bottom).
  The horizontal beam walk is denoted with $y$, and the vertical beam walk with $z$.}
\label{fig:TTLspotpositions_B}
\end{figure}

For completeness, we show in Fig.~\ref{fig:TTLspotpositions_B} the weighted spot positions measured during the LPF mission for B-diodes of the xR- and the x12-interferometer.
The result is very similar to the spot positions measured for the corresponding A-diodes (Fig.~\ref{fig:TTLspotpositions}).
Thus, the small difference in beam path does not significantly affect the beams' points of detection, and both readouts could equivalently be used for the analysis.

\bibliographystyle{unsrt}
\def\bibsection{\section*{References}} 
\bibliography{References.bib}
\end{document}